\documentclass[12pt,english,superscriptaddress,nofootinbib,preprintnumbers, aps]{revtex4}
\usepackage{color}
\usepackage{amsmath}
\usepackage{amsthm}
\usepackage{graphicx}
\usepackage{amssymb}
\usepackage{dsfont}
\usepackage{color}
\usepackage{soul}
\usepackage{esint}
\usepackage[hyperfootnotes=true,unicode=true, 
 bookmarks=true,bookmarksnumbered=false,bookmarksopen=false,
 breaklinks=false,pdfborder={0 0 1},backref=false,colorlinks=true]
 {hyperref}
 \usepackage{float}
 \usepackage[markup=nocolor]{changes}

\usepackage[makeroom]{cancel}

\usepackage{etoolbox}

\makeatletter
\@ifundefined{textcolor}{}
{%
 \definecolor{BLACK}{gray}{0}
 \definecolor{WHITE}{gray}{1}
 \definecolor{RED}{rgb}{1,0,0}
 \definecolor{GREEN}{rgb}{0,1,0}
 \definecolor{BLUE}{rgb}{0,0,1}
 \definecolor{CYAN}{cmyk}{1,0,0,0}
 \definecolor{MAGENTA}{cmyk}{0,1,0,0}
 \definecolor{YELLOW}{cmyk}{0,0,1,0}
 }

\usepackage{hyperref}
\hypersetup{
    colorlinks,%
    citecolor=blue,%
    filecolor=blue,%
    linkcolor=blue,%
    urlcolor=blue
}
\patchcmd{\section}
  {\centering}
  {\raggedright}
  {}
  {}
\patchcmd{\subsection}
  {\centering}
  {\raggedright}
  {}
  {}

\makeatother

\begin{document}

%\title{Coherent State Expectation-Values in LQG \\ I. Isotropic, Flat Cosmology}

\title{
Cosmological Coherent State Expectation Values in LQG \\ I. Isotropic Kinematics}

\author{Andrea Dapor}
\email{andrea.dapor@gravity.fau.de}
\affiliation{Institute for Quantum Gravity, Friedrich-Alexander University Erlangen-N\"urnberg, Staudstra\ss e 7, 91058 Erlangen, Germany}

\author{Klaus Liegener}
\email{klaus.liegener@gravity.fau.de}
\affiliation{Institute for Quantum Gravity, Friedrich-Alexander University Erlangen-N\"urnberg, Staudstra\ss e 7, 91058 Erlangen, Germany}

\date{\today{}}

\begin{abstract}

\fontfamily{lmss}\selectfont{This is the first paper of a series dedicated to LQG coherent states and cosmology. The concept is based on the effective dynamics program of Loop Quantum Cosmology, where the classical dynamics generated by the expectation value of the Hamiltonian on semiclassical states is found to be in agreement with the quantum evolution of such states. We ask the question of whether this expectation value agrees with the one obtained in the full theory. The answer is in the negative, \cite{DaporKlaus}. This series of papers is dedicated to detailing the computations that lead to that surprising result. In the current paper, we construct the family of coherent states in LQG which represent flat ($k = 0$) Robertson-Walker spacetimes, and present the tools needed to compute expectation values of polynomial operators in holonomy and flux on such states. These tools will be applied to the LQG Hamiltonian operator (in Thiemann regularization) in the second paper of the series. The third paper will present an extension to $k \neq 0$ cosmologies and a comparison with alternative regularizations of the Hamiltonian.}
\end{abstract}

\maketitle
\section{Introduction}
\label{s1}
\numberwithin{equation}{section}
Virtually every area of quantum physics is, at least partially, concerned with coherent states. This is due to their relevance in relation to the classical limit of quantum theories. Indeed, a minimum requirement for a family of coherent states is that each state is labelled by a point in the phase space of the theory. This is then interpreted as the quantum state that represents the closest approximation to the classical configuration of that phase space point.
\\
In the context of Loop Quantum Gravity (LQG) \cite{Rov04,AL04,Thi07} -- a non-perturbative quantization of General Relativity (GR) in its 3+1 ADM formulation -- coherent states have been constructed based on the concept of weave states \cite{weave} and heat kernel techniques for compact groups \cite{Hall1,Hall2}. The resulting states have been called ``complexifier coherent states'' \cite{Winkler1,Winkler2,Winkler3,SahThiWin,ThiemanComplex} or ``gauge coherent states'' (owing to the fact that they are covariant under the $SU(2)$ gauge group of LQG). Their properties and their relations to other semiclassical states have been thoroughly investigated \cite{SahThi1,SahThi2,BahThi1,BahThi2,MagliaroPerini}. It is important to notice that gauge coherent states are based on a single graph (or a sum over countably many graphs).\footnote
{
If we naively take a linear combination of such fixed-graph coherent states on the label-set of graphs, we discover that there is no damping factor fast enough to make the norm of the state finite. This problem is due to the non-separability of the kinematical Hilbert space of LQG, and might be solved at the physical level. However, a definition of coherent states even at the diffeomorphism-invariant level is still missing, though proposals exist \cite{ALMMT}.
}
This means that they cannot describe continuous geometries. Nevertheless, if one limits their interest to the Hilbert space of a fixed graph (which we will do), each state in the family represents a (discrete) classical 3-geometry. This is perfectly fine for most applications, and a theoretical justification for working on a fixed graph was given in the framework of Algebraic Quantum Gravity (AQG) \cite{AQG1,AQG2,AQG3}. In this formulation of LQG, one works with an abstract graph (that is, not embedded in a 3-manifold, which has the added benefit of simplifying the implementation of diffeomorphism symmetry), and hence only excitations along already existing edges are allowed.\footnote
{
As a consequence, we can only consider non-graph-changing operators. This is particularly important for the Hamiltonian operator, as one must abandon the original regularization \cite{Thi96_1,Thi96_2} and consider instead a non-graph-changing one \cite{AQG2}.
}
\\
In this series of papers, we wish to consider the applications of gauge coherent states to the description of classical cosmological spacetimes. Due to the homogeneity that characterizes such geometries, the choice of the fixed graph is that of a regular lattice, which we take to be cubic (i.e., each vertex is 6-valent) so that it can be oriented along a fiducial system of coordinates of the 3-geometry. This is the same starting point of Loop Quantum Cosmology (LQC) \cite{LQC1,LQC2,LQC3,LQC4}, which is a LQG-inspired quantization of homogeneous spacetimes: the hope is therefore to shed light on the relation between the full theory of LQG and the mini-superspace models described by LQC. Particularly, we shall be interested in the so-called ``effective dynamics'': in LQC it was numerically shown \cite{AshPawSin} that the expectation value of the Hamiltonian $\hat H_{LQC}$ on gaussian states $\psi_{(c,p)}$ plays the role of effective Hamiltonian. By this, we mean that the Hamiltonian flow produced by $H_{\text{eff}}(c,p) := \langle \psi_{(c,p)} | \hat H_{LQC} | \psi_{(c,p)} \rangle$ on the $(c,p)$-phase space of cosmology coincides with the trajectories followed by the peak of gaussian states when they are evolved via $\hat H_{LQC}$:
\begin{align}
\langle \psi_{(c,p)} | e^{i \phi \hat H_{LQC}} \hat F e^{-i \phi \hat H_{LQC}} | \psi_{(c,p)} \rangle = e^{\{\cdot, H_{\text{eff}}(c,p)\}} \langle \psi_{(c,p)} | \hat F | \psi_{(c,p)} \rangle + \mathcal O(s)
\end{align}
for $\hat F$ an operator on the Hilbert space of LQC, $\phi$ some matter degree of freedom playing the role of physical time (e.g., a massless scalar field), and $s$ the spread of state $\psi_{(c,p)}$. The question of whether this feature lifts to full LQG is still open, but we shall nevertheless refer to the expectation value of the full Hamiltonian on our coherent sates as ``effective Hamiltonian'', and we ask whether the dynamics it generates on the phase space coincides with that of LQC. The answer has been found affirmative in toy models such as Quantum Reduced Loop Gravity \cite{QRLG1,QRLG2} but for full LQG it turns out to be in the negative, and the result was discussed in the short paper \cite{DaporKlaus}. The current series of papers is a detailed presentation of the techniques developed and used to derive this result.
\\
In this first paper, we identify the classical degrees of freedom corresponding to a $k=0$ Robertson-Walker spacetime and propose the subfamily of gauge coherent states that describes such classical spacetimes. Then we develop the techniques needed to compute expectation values (and, ultimately, matrix elements) of operators on this subfamily. Applying this technology to the elementary operators (i.e., holonomies and fluxes) and to their dispersions (spreads), we confirm that each of these states is peaked on the classical Robertson-Walker geometry that labels it. We finally apply the techniques to the volume operator, computing its expectation value to next-to-leading order in the semiclassicality parameter that controls the spread of the states: we find that the leading order is in agreement with the classical volume of a cubic cell in flat Robertson-Walker spacetime.
\\
\\
The article is organized as follows.
\\
In section \ref{s2} we briefly review LQG, with particular attention to the quantization of the scalar constraint, as it is related to the Hamiltonian in the full theory. We also discuss the the basics of gauge coherent states.
\\
In section \ref{s3} the subfamily of cosmological coherent states will be presented. We also prove certain relaions satisfied by these states, which will drastically simplify the computations in the next sections.
\\
In section \ref{s4} we compute the expectation values of monomials in the fundamental operators (holonomy and flux), as these are the basic building blocks for the interesting geometric operators.
\\
In section \ref{s5} the tools are put into action, as we compute the expectation values and dispersions of the fundamental operators, showing that the states are indeed peaked on homogeneous isotropic geometries. We also compute the expectation value of what we call the ``Giesel-Thiemann Volume Operator'' which, as it was shown in \cite{AQG3}, coincides with the Ashtekar-Lewandowski Volume Operator up to a desired order in $\hbar$.
\\
In section \ref{s6} we summarize our results and briefly comment on further research.
\\
Finally, a word on the appendices: appendix \ref{recovering-gauge} deals with the question of gauge-invariance as far as our cosmological coherent states are concerned; appendix \ref{basic-properties} collects some properties of $SU(2)$ irreducible representations and recoupling theory (used in computations throughout the text); appendix \ref{hol-int} contains the explicit computation of certain integrals relevant for expectation values involving the holonomy operator.

\section{Review of LQG and gauge coherent states}
\label{s2}
This section is merely a recap of known results in order to clarify the notation used in this article. The experienced reader may jump directly to section \ref{s3}.
		\subsection{Ashtekar-Barbero Variables}
We cast general relativity into its hamiltonian formulation, by splitting our four-dimensional manifold $\mathcal{M}=\mathbb{R}\times \sigma$, where $\sigma$ is a smooth 3-manifold which admits Riemannian metrics. On $\Sigma$ we can define triad fields $e^a_I$, as well as co-triads $e^I_a$ where $a,b,c...\in\{1,2,3\}$ denote tensorial indices and $I,J,K..\in\{1,2,3\}$ can be thought of as $\mathfrak{su}(2)$ algebra indices. These triads are subject to the condition that the 3-metric can be derived from them via $q_{ab}=e^I_ae^J_b \delta_{IJ}$ and that $e^a_Ie^I_b=\delta^a_b$, $e^I_ae^a_J=\delta^I_J$. We then introduce the Ashtekar-Barbero variables \cite{AB-variables1,AB-variables2,AB-variables3}, i.e., the lie-algebra-valued $\text{SU}(2)$-connection $A_a=A_a^I\tau_I$ and the densitized triad $E^a_I$:
\begin{align}
A^I_a (x) := \Gamma^I_a(x)+\beta K_{ab}(x)e^b_J\delta^{JI},\hspace{20pt} E^a_I (x) := \mid \det(e)\mid e^a_I
\end{align}
with $\Gamma^I_a$ the complex-valued spin-connection of $e^I_a$, $K_{ab}$ the extrinsic curvature of $\sigma$, $\det(e):=\det(\{e^J_c\}^J_c)$ and $\beta\in \mathbb{R}$ the Immirzi-parameter. By $\tau_I = -i\sigma_I$ we denote the imaginary Pauli matrices, which in cartesian coordinates $(I=1,2,3)$ are given by
\begin{align}
\tau_1=-\left(\begin{array}{ccc}
0 & i\\
i & 0
\end{array}\right), \hspace{15pt}
\tau_2=-\left(\begin{array}{ccc}
0 & 1\\
-1 & 0
\end{array}\right), \hspace{15pt}
\tau_3=-\left(\begin{array}{ccc}
i & 0\\
0 & -i
\end{array}\right), \hspace{15pt}
\end{align}
They are the generators of the lie algebra $\mathfrak{su}(2)$ and fulfill $\text{Tr}(\tau_I\tau_J)=-2\delta_{IJ}$ and $[\tau_I,\tau_J]=2\epsilon_{IJ}^{\hspace{8pt}K}\tau_K$.
\\
The Ashtekar-Barbero variables form a canonical pair $(A^I_a,E^a_I)$, i.e., we find for their Poisson-bracket that $\{A^I_a (x), E^b_J(y)\}=\delta^b_a \delta^I_J\delta^{(3)}(x,y)$. They uniquely identify a physical geometry if one imposes the {\it Gauss constraint}
\begin{align}
G_J:= \partial_a E^a_J + \epsilon_{JKL}\delta^{LM} A^K_a E^a_M
\end{align}
and the standard constraints of GR: the {\it vector constraint} (smeared with an arbitrary shift-function $N^a$)
\begin{align}
\bar{C}_{Diff}[\bar{N}]=\int dx^3 N^a (x) F_{ab}(x)^IE_I^b(x) 
\end{align}
and the {\it scalar constraint} (smeared with an arbitrary lapse function $N$)
\begin{align}
C [N]&=\int dx^3 N \left( F^I_{ab}-(\beta^2+1)\epsilon_{I}^{\hspace{2pt}LM}K_{ac}e^c_LK_{bd}e^d_M\right)\frac{\epsilon^{IJK}E^a_JE^b_K}{\sqrt{\det(q)}}
\end{align}
where $F_{ab}^I:= 2\partial_{[a}A^I_{b]}+\epsilon_{IJK}A^J_aA^K_b$ is the curvature of the connection. For the purpose of quantization, one uses the Thiemann identities \cite{Thi96_1,Thi96_2}
\begin{align}
\{V,A^J_a\}=e^J_a=\frac{1}{2}sgn(\det(e))\epsilon^{JKL}\epsilon_{abc}\frac{E^b_KE^c_L}{\sqrt{\det(q)}},\hspace{20pt}\{V,C_E[1]\}=\beta^2\int dx^3 K_{ab}e^b_I\delta^{IJ} E^a_J
\end{align}
to rewrite the scalar constraint as
\begin{align} \label{ScalarConstraint}
C [N]=\int dx^3 N \left(F^I_{ab}\delta_{IJ} -(\beta^2+1)\epsilon_{JMN} \{\{V,C_E[1]\},A^M_a\}\{\{V,C_E[1]\},A^N_b\} \right) \epsilon^{abc}\{V,A^J_c\}
\end{align}
where $C_E[N] := \int d^3x N F^I_{ab}\delta_{IJ}\epsilon^{abc}\{V,A^J_c\}$ is called the {\it euclidean Hamiltonian} and $V :=\int dx^3 \sqrt{\det(q)}$ is the total volume of the manifold.
		\subsection{Quantization}
As we are now in the situation of a gauge theory, we can quantize the phase space variables and the constraints following Dirac procedure. We smear the connection $A_a(x)$ along any curve $e$ in the manifold to obtain the {\it holonomy} $h(e)\in\text{SU}(2)$, i.e., the path-ordered exponential of the connection along $e$. We similarly smear the densitized triad $E^a$ against any 2-dimensional surface $S$ to obtain the {\it flux} $E(S)\in\mathfrak{su}(2)$. For a curve $e$ and a surface $S$ we thus define
\begin{align}
h(e):=\mathcal{P}\exp\left(\int_0^1 dt A_a (c(t))\dot{c}^a(t)\right),\hspace{10pt} E^I(S):=\int_S \epsilon_{abc} dx^a\wedge dx^b E^c_J\delta^{IJ} 
\end{align}
The set of $(h,E)$ along {\it all} curves and surfaces constitutes the {\it holonomy-flux algebra}, which one uses to define the Hilbert space by GNS construction. If one fixes finitely many oriented curves, the union of them forms a graph, $\gamma=\bigcup_l e_l$. We call $e_l\in E(\gamma)$ an edge (or link) and any intersection $v \in V(\gamma)$ of two edges a vertex (or node). One can then associate a Hilbert space to $\gamma$ by considering the tensor product of square integrable functions on each edge, $\mathcal{H}:=\otimes_{e\in E(\gamma)} \mathcal{H}_e$ with $\mathcal{H}_e=L_2(\text{SU}(2),d\mu_H)$, $d\mu_H$ being the unique Haar measure on $\text{SU}(2)$. The elements $F_{\gamma}\in \mathcal{H}$ are called {\it cylindrical functions}. The holonomies get promoted to bounded, unitary multiplication operators: for $f_e\in\mathcal{H}_e$ it is
\begin{align}
\hat{h}_{mn}(e)f_e(g):=D^{(\frac{1}{2})}_{mn}(g)f_e(g)
\end{align} 
where $D^{(\frac{1}{2})}_{mn}(g)$ is the Wigner-matrix of group element $g$ in the defining irreducible representation of $\text{SU}(2)$ corresponding to spin-1/2 \cite{Brink-Satchler}. The Peter-Weyl Theorem ensures that they form an orthogonal basis, hence any function in $\mathcal H_e$ can be written as $f_e(g_e)=\sum_{j}\sum_{-j\leq m,n\leq j} c_{jmn} D^{(j)}_{mn}(g_e)$, where $j\in \mathbb{N}/2$ (sums over magnetic indices $m,n,...$ will be suppressed in the following). The scalar product is given in the $L_2$ sense:
\begin{align}
\langle F_{\gamma},F_{\gamma'}'\rangle=\delta_{\gamma,\gamma'} \prod_{e\in\gamma} \int d\mu_{H}(g_e) \overline{f_e(g_e)} f_e' (g_e) \\
\int d\mu_H(g) \overline{D^{(j)}_{mn}(g)} D^{(j')}_{m'n'}(g)=\frac{1}{d_j}\delta_{jj'}\delta_{mm'}\delta_{nn'}\label{orthogInt}
\end{align}
where the dimension of spin-j $\text{SU}(2)$-irrep is $d_j=2j+1$. Similiarly, the fluxes become essentially self-adjoint derivation operators:
\begin{align} \label{fluxy}
\hat{E}^K(S)f_e(g):= -\frac{i\hbar\kappa\beta}{4} \sigma(e\cap S) f_{e_1}(g_{e_1}) R^K(e_{2})f_{e_2}(g_{e_2})
\end{align}
where $\sigma(e\cap S)\in\{0,\pm 1\}$ (depending on if edge and surface meet non-transversally or under the same/opposite orientation respectively), $e=e_1 \circ e_2$ such that $s_e=e\cap S$ is the starting point of edge $e_2$ and $g=g_{e_1}g_{e_2}$ (which makes the splitting unique). Finally, the {\it right-invariant vector field} $R^K(e)$ is defined together with the {\it left-invariant vector field} $L^K(e)$ as
\begin{align}
R^K(e)f_e(g):=\left.\frac{d}{ds}\right|_{s=0} f_e(e^{s\tau_K}g),\hspace{20pt} L^K(e)f_e(g):=\left.\frac{d}{ds}\right|_{s=0} f_e(ge^{s\tau_K})
\end{align}
In particular, the action of $R^K$ on the basis element is given by
\begin{align} \label{R-and-D-prime}
R^KD^{(j)}_{mn}(g) = {D'}^{(j)}_{m\mu}(\tau_K)D^{(j)}_{\mu n}(g), \ \ \ {D'}^{(j)}_{mn}(\tau_K) = 2i\sqrt{j(j+1)d_j}(-1)^{j+n}\left(\begin{array}{ccc}
j & 1 & j\\
n & K & -m
\end{array}\right)
\end{align}
as is shown in appendix \ref{basic-properties}. This concludes the description on the {\it kinematical Hilbert space} of LQG.
\\
It remains to incorporate the constraints $G_J, \bar{C}_{Diff}[\bar{N}], C[N]$. The Gauss constraint $G_J$ is easily incorporated by the fact that it is the generator of $\text{SU}(2)$-rotations, hence its solutions are states of $\mathcal{H}$ which are invariant under $\text{SU}(2)$. These can be obtained by group averaging: let $U_G[g]$ be the operator that generates a local $g(x)\in\text{SU}(2)$ transformation and $F_{\gamma}(\{g\})=\prod_{e\in\gamma} f_e(g_e)$; then the corresponding gauge-invariant function is
\begin{align}\label{Group-Averaging}
F_{\gamma}^G(g)= \int D[\{h\}] U_G[\{h\}] F_{\gamma}(\{g\}) := \left( \prod_{v\in V(\gamma)} \int d\mu_H (h_v)\right) \prod_{e\in\gamma} f_e(h_{s_e} gh_{t_e}^{-1})
\end{align}
where $v$ runs through all vertices in $\gamma$, and $s_e, t_e$ denote the vertex at the beginning/end of edge $e$ respectively.\\
The vector constraint $\bar{C}_{Diff}[\bar{N}]$ generates diffeormorphisms of the spatial manifold $\sigma$, and therefore cannot be implemented as an infinitesimal operator due to the action of the diffeomorphism group $\text{Diff}(\sigma)$ not being strongly continuous. Nevertheless, diffeomorphism-invariance can still be implemented via finite diffeomorphisms $\varphi\in\text{Diff}(\sigma)$. For this purpose, in this paper we adopt the idea developed in the context of AQG \cite{AQG1}, where one considers {\it abstract graphs}, that is, graphs which ``forget'' about their embedding in $\sigma$. We will talk about what this explicitly means in section \ref{s3}, when we choose the states with respect to which we compute expectation values.
\\
Finally, let us consider the implementation of the scalar constraint in the quantum theory. Following the strategy of \cite{Thi96_1,Thi96_2}, we rewrite (\ref{ScalarConstraint}) in terms of holonomies and then promote every term to an operator. For the volume -- which was pivotal for using the Thiemann-identities -- this leads to the Ashtekar-Lewandowski volume operator \cite{volume1,volume2}:
\begin{align}
\hat{V}(\sigma) F_{\gamma}(\{g\}) = \ & \frac{(\beta\hbar\kappa)^{3/2}}{2^{5}\sqrt{3}}\sum_{v\in V(\gamma)}\hat{V}_v F_{\gamma}(\{g\}) =\frac{(\beta\hbar\kappa)^{3/2}}{2^{5}\sqrt{3}}\sum_{v\in V(\gamma)}\sqrt{\mid \hat{Q}_v \mid}F_{\gamma}(\{g\}), \label{ALvolume}
\\ 
\hat{Q}_v := \ & i\sum_{e\cap e'\cap e'' =v } \epsilon(e,e',e'')\epsilon_{IJK}R^I(e)R^J(e')R^K(e'') \label{Q-oper}
\end{align}
with $\epsilon(e, e', e'') := sgn(\det(\dot{e},\dot{e}',\dot{e}''))$ and all edges outgoing at the vertex $v$. (In case of an $e$ being ingoing, one simply replaces $R^K(e)\rightarrow L^K(e)$.) Since the square-root is understood in the sense of the spectral theorem, knowledge of the full spectrum of $\hat{Q}_v$ is required before we can say how $\hat{V}_v$ acts on any state. Unfortunately, despite a lot of research has been done \cite{volume3,volume4,volume5} on the spectrum of (\ref{ALvolume}), a general formula for its eigenstates is still unknown.
\\
Among the various choices of regularization proposed for the scalar constraint, we will use the framework first developed in AQG \cite{AQG2}, where one chooses the scalar constraint to act in a {\it non-graph-changing} way, i.e., one regularizes the curvature of the Ashtekar connection by $F_{ab}(x)\dot{e}^a\dot{e}^b=[h(\square_{ee'})-h(\square_{ee'})^{\dagger}]/2\epsilon^2+\mathcal{O}(\epsilon)$ where $\square_{ee'}$ denotes a small loop starting at $x$ along $e$ and returning along $e'$. Then, we choose for the action of the loop-holonomy the operator $\hat h(\square_{ee'})$, which starts at a vertex $v$ of the graph and goes along already existing, excited edges of $F_{\gamma}$. Thus, the total operator in its symmetrized version looks as follows:
\begin{align}
\hat{C}[N] = \frac{1}{2}\left(\hat{C}_E+\hat{C}_E^{\dagger}\right)-\frac{\beta^2+1}{2}\left(\hat{C}_L+\hat{C}_L^{\dagger}\right)
\end{align}
where
\begin{align}
\hat{C}_E[N]:=&\frac{32}{3i\kappa^2\hbar\beta}\sum_{v\in V(\gamma)}\frac{N_v}{20}\sum_{e\cap e' \cap e''=v}\epsilon(e,e',e'')\frac{1}{2}\times\nonumber\\
&\hspace{30pt}\times\text{Tr}\left((\hat{h}(\square_{ee'})-\hat{h}(\square_{ee'})^{\dagger})\hat{h}(e'')\left[\hat{h}(e'')^{\dagger},\hat{V}_v\right]\right)\label{EuclHam}
\\
\hat{C}_L[N]:=&\frac{128}{3i\kappa^4\hbar^5\beta^5}\sum_{v\in V(\gamma)}\frac{N_v}{20}\sum_{e\cap e'\cap e''=v}\epsilon(e,e',e'')\times\nonumber\\
&\hspace{30pt}\times\text{Tr}\left(\hat{h}(e)\left[\hat{h}(e)^{\dagger},[\hat{C}_E[1],\hat{V}_v]\right]
\hat{h}(e')\left[\hat{h}(e')^{\dagger},[\hat{C}_E[1],\hat{V}_v]\right]
\hat{h}(e'')\left[\hat{h}(e'')^{\dagger},\hat{V}_v\right]\label{LorHam}
\right)
\end{align}
and $N_v$ is the value of lapse function $N$ at $v \in \sigma$.
		\subsection{Deparametrization with Gaussian Dust}
Instead of dealing with vacuum GR, where one has to solve the scalar constraint $C[N]$, one can construct observables by adding matter to the theory and trying to find local coordinates such that the constraint acquires the form $C = P + H$ in terms of the conjugated momentum $P$ to the matter degree of freedom. If this form is achieved, one speaks of ``relational observables'' and ``deparametrization'' \cite{dep1,dep2,dep3,dep4,dep5,dep6,dep7}: the function $H$ becomes a physical, conserved Hamiltonian density which drives the physical evolution of the observables with respect to the matter degree of freedom (which therefore plays the role of physical clock, $\tau$). While not all types of matter allow for this decomposition, a good choice is Gaussian dust: in the framework of Torre and Kucha$\check{\text{r}}$ \cite{dep1}, the Lagrangian added to the Einstein-Hilbert action describing Gaussian dust is
\begin{align}\label{TorreKuchar}
\mathcal{L}_{GD}=- \sqrt{\mid \det(g)\mid}\left(\frac{\varrho}{2}(g^{\mu\nu}T_{,\mu}T_{,\nu}+1)+g^{\mu\nu}T_{,\mu}(W_jS^j_{,\nu})\right)
\end{align}
with the field $\varrho$ having dimension $[\text{cm}^{-4}]$, the fields $T,S^j$ having dimension $[\text{cm}]$, and $W_j$ being dimensionless. Performing Legendre transformation, one can show that the time-evolution of an observable ${O}_F$ (associaed with phase space function $F$) is encoded as the Schr\"odinger-like equation $d{O}_F(\tau)/d\tau=\{H, {O}_F(\tau)\}$, where
\begin{align}
H = C[1] = \int dx^3 C(x)
\end{align}
is for this reason called the {\it true Hamiltonian}. We see that $C$ it is not longer a constraint whose vanishing must be imposed, but in fact it generates time-translations. Thus, if one takes this viewpoing, the quantum scalar contraint presented above is understood as the quantum operator producing the dynamics of geometric degrees of freedom wrt the classical observer provided by the dust.
		\subsection{Gauge Coherent States}
We have now at our disposal a physical Hilbert space on the fixed graph $\gamma$, and have an understanding of what we mean by dynamics. But while any state $F \in \mathcal H$ can be considered, in this work we will focus on a subset of the {\it gauge coherent state} family. Let us therefore briefly review the general definition and properties of this family.
\\
Following Hall \cite{Hall1, Hall2}, one constructs a coherent state $\psi^t_{e,h_e^{\mathbb{C}}}$ for every edge $e$ of the graph, and glues them together in a cylindrically-consistent way obtaining $\Psi^t_{\gamma,\{h^{\mathbb{C}}\}}(\{g\}):=\prod_{e\in E(\gamma)}\psi^t_{e,h^{\mathbb{C}}_e}(g_e)$. To construct $\psi^t_{e,h_e^{\mathbb{C}}}$ one uses a complex polarization of the classical phase space, i.e., a unitary map $(A,E) \mapsto A^{\mathbb{C}}$ that expresses the complex connection as a function of the real phase space. For example, the left-polar decomposition prescribes
\begin{align} \label{polar-decomposition}
h_e^{\mathbb{C}}:=\exp\left(-\frac{it}{\hbar\kappa\beta} \tau_J E^J(S_e)\right) h(e) \in SL(2,\mathbb C)
\end{align}
where $h(e)$ is the classical holonomy along edge $e$ and $E^J(S_e)$ is the classical flux across the open surface $S_e$ manually assigned to each $e\in E(\gamma)$ in such a way that (1) all $S_e$ are mutually non-intersecting, (2) only $e$ intersects $S_e$ and the intersection is transversal and consists of only one point, (3) both $S_e$ and $e$ carry the same orientation. The dimensionless quantity $t:=\hbar\kappa/a^2 > 0$ is called the {\it semiclassicality parameter}, with $a$ being a length scale that the theory should provide.\footnote
{
As we will see later, $t$ controls the spread in holonomy and flux of the coherent state: The smaller $t$, the smaller the relative dispersions of $h$ and $E$. It has been therefore argued \cite{AQG3} that the natural choice for $a^2$ in a vacuum gravity context is the inverse of cosmological constant, $a^2 = 1/\Lambda$. Using $\kappa = 16\pi G/c^3$, one then finds $t \sim 10^{-120}$.
}
\\
To construct the coherent state in $\mathcal H_e$ peaked on $h_e^{\mathbb{C}} \in SL(2,\mathbb{C})$, one first chooses a {\it complexifier} $\hat{C}_t$ and exponentiates it: this gives rise to the coherent state transform, which for the choice of heat kernel complexifier \cite{Winkler1} reads
\begin{align}
\hat{W}_t := e^{-\frac{1}{\hbar}\hat{C}_t} = e^{\frac{t}{8}\delta_{IJ}R^I(e)R^J(e)}
\end{align}
The {\it (gauge-variant) coherent state} is now obtained by applying $\hat{W}_t$ to the delta-function on $\text{SU}(2)$, $\delta_{h'}$, and continuing analytically the result to $h' \rightarrow h_e^{\mathbb{C}}$:
\begin{align}\label{gauge-variant coh}
\psi^t_{e,h^{\mathbb{C}}_e}(g):=\left(\hat{W}_t\delta_{h'}(g)\right)_{h' \rightarrow h^{\mathbb{C}}_e} =\sum_{j} d_j e^{-\frac{t}{2}j(j+1)}\text{Tr}^{(j)}((h_e^{\mathbb{C}})^\dag g)
\end{align}
where $\text{Tr}^{(j)}(.)$ denotes the trace in the spin-$j$ irrep and the explicit expression $\delta_{h}(g)=\sum_{j}d_j\text{Tr}^{(j)}(hg^{-1})$ has been used.
\\
As was shown in \cite{Winkler2}, these coherent states fulfill a number of useful properties:
\begin{itemize}
\item[(1)] {\it Eigenstates of an annihilation operator}. By defining $\hat{a}(e) := e^{-\frac{1}{\hbar}\hat{C}_{t}}\hat{h}(e)e^{\frac{1}{\hbar}\hat{C}_t}=e^{\frac{3t}{8}}e^{-i\tau_I\hat{E}^I(S_e)/2}\hat{h}(e)$ (where the action of the last exponential has to be understood via Nelson's analytic vector theorem), one finds that the coherent states are simultanous eigenstates for each one:
\begin{align}
\hat{a}_{mn}(e)\Psi^t_{\gamma,\{h^{\mathbb{C}}\}}= [h_e]_{mn}\Psi^t_{\gamma,\{h^{\mathbb{C}}\}}
\end{align}
\item[(2)] {\it Overcompleteness relation}. By considering the measure $d\nu_t(e^{i\tau_J p_J /2}h_e):=d\mu_H(h_e)[\frac{2\sqrt{2}e^{-t/4}}{(2\pi t)^{3/2}}\frac{\sinh(\sqrt{p^2})}{\sqrt{p^2}}e^{-p^2/t}dp^3]$ on $\text{SL}(2,\mathbb{C})$, one can show that
\begin{align}
\int_{\text{SL}(2,\mathbb{C})}d\nu_t(h) \ \psi^t_{e,h_e^{\mathbb{C}}} \ \langle \psi^t_{e,h_e^{\mathbb{C}}}, \cdot \rangle = \mathds{1}_{\mathcal{H}_e}
\end{align}
\item[(3)] {\it Peakedness in holonomy and electric flux}. For all $h,h'\in\text{SL}(2,\mathbb{C})$ there exists a positive function $K_t(h,h')$ decaying exponentially fast as $t\rightarrow 0$ for $h\neq h'$ and such that
\begin{align}
\mid\langle \psi^t_h,\psi^t_{h'}\rangle\mid^2\leq K_t(h,h') ||\psi^t_h||^2||\psi^t_{h'}||^2
\end{align}
Moreover, for holonomies and fluxes one finds
\begin{align}
\langle \psi^t_h,\hat{h}_{mn}(e)\psi^t_{h'}\rangle = \ & h_{mn}(e) \langle \psi^t_h,\psi^t_{h'}\rangle + \mathcal{O}(t)
\\
\langle \psi^t_h,\hat{E}^J(S_e)\psi^t_{h'}\rangle = \ & E^J(S_e) \langle \psi^t_h,\psi^t_{h'}\rangle + \mathcal{O}(t)
\end{align}
\end{itemize}
Most importantly for our purposes, the advantage of using these coherent states is that the evaluation of expectation values of operators involving the Ashtekar-Lewandowski volume (\ref{ALvolume}) can be drastically simplified. Indeed, given a ``good'' coherent state (i.e., one which is peaked at each edge on $|h_{mn}(e)| \gg t$, $|E^J(S_e)| \gg t$), it was shown in \cite{AQG3} that for every polynomial operator $P(\hat{V}_v,\hat{h})$ the following relation holds:
\begin{align}\label{Replacement}
\langle \Psi^t_{\gamma, \{h^{\mathbb{C}}\}}, P(\hat{V}_v,\hat{h})\Psi^t_{\gamma, \{h^{\mathbb{C}}\}} \rangle = \langle \Psi^t_{\gamma, \{h^{\mathbb{C}}\}}, P(\hat{V}_{k,v}^{GT},\hat{h})\Psi^t_{\gamma, \{h^{\mathbb{C}}\}} \rangle + \mathcal{O}(t^{k+1})
\end{align}
where we refer to $\hat{V}^{GT}_v$ as the {\it $k$-th Giesel-Thiemann volume operator}. This is explicitly given by
\begin{align}\label{GTvolume}
\hat{V}^{GT}_{k,v}:= \langle \hat{Q}_v \rangle^{1/2}\left[\mathds{1}_{\mathcal{H}} + \sum_{n=1}^{2k+1}\frac{(-1)^{n}}{n!} \left(0-\frac{1}{4}\right) \left(1-\frac{1}{4}\right) ... \left(n-1-\frac{1}{4}\right) \left(\frac{\hat{Q}_v^2}{\langle \hat{Q}_v \rangle^{2}}-\mathds{1}_{\mathcal{H}}\right)^n
\right]
\end{align}
where $\hat{Q}_v$ is as in (\ref{Q-oper}) and we used the shorthand notation $\langle \hat Q_v \rangle := \langle \Psi^t_{\gamma, \{h^{\mathbb{C}}\}}, \hat Q_v \Psi^t_{\gamma, \{h^{\mathbb{C}}\}} \rangle$.\footnote
{
We observe that operator $\hat{V}^{GT}_{k,v}$ depends explicitly on the coherent state $\Psi^t_{\gamma, \{h^{\mathbb{C}}\}}$ which appears in (\ref{Replacement}), and it therefore makes sense only in the context of equation (\ref{Replacement}).
}
This fact enables us to compute the approximated expectation value (on these coherent states) of any polynomial operator involving Ashtekar-Lewandowski volume, retaining control on the error we make in terms of powers of the semiclassicality parameter $t$.
\section{Cosmological gauge coherent states for LQG}
\label{s3}
In this section we will focus on a subfamily of the coherent states $\Psi^t_{\gamma, \{h^{\mathbb{C}}\}}$, which we claim to be suited to describe flat Robertson-Walker geometries (this claim will be substantiated in the next two sections) at a given instant, i.e., on the spatial manifold $\sigma$. The question of whether these states are actually stable under the dynamics is still open, and will not be addressed here.
		\subsection{Choice of States}
We introduce an infrared cutoff by restricting the spatial manifold $\sigma$ to a compact submanifold, $\sigma_R$, which we equip with the topology of a 3-Torus, that is, periodic boundary conditions. With respect to a fiducial metric $\eta$ we identify $R$ as the coordinate length of the Torus, which in principle allows us to remove the cutoff by sending $R \rightarrow \infty$. Thus, we are interested in a fixed graph $\gamma$, which is chosen to be a compact subset of the cubic lattice $\mathbb{Z}^3$ embedded in $\sigma_R$. As such, we shall only consider a subalgebra of the holonomy-flux algebra: the holonomies along the edges of $\gamma$ and the fluxes across the surfaces of the dual cell-complex. To be precise the algebra of the resulting operators reads:
\begin{align}\label{OperatorAlgebra}
\begin{array}{c}
[\hat{h}_{ab}(e),\hat{h}_{cd}(e')]=0,\hspace{15pt}\hspace{15pt}[R^K(e),R^L(e')]=\delta_{ee'}\epsilon^{KL}_{\hspace{8pt}M} R^M(e)
\\
\\
{[}R^K(e),\hat{h}_{ab}(e'){]}=\delta_{ee'}{D'}^{(\frac{1}{2})}_{ac}(\tau_K)\hat{h}_{cb}(e)
\end{array}
\end{align}
with $D'^{(\frac{1}{2})}_{ab}(\tau_K)$ as defined in (\ref{R-and-D-prime}).
\\
The three directions of the lattice can be chosen adapted to the fiducial metric $\eta$, so that the coordinate length of a side of the lattice is $R$. On the other hand, due to $\sigma_R$ being compact, $\gamma$ has a finite number of vertices, $\mathcal{N}^3$. Assuming the lattice to be regular wrt to $\eta$, we therefore find that the coordinate distance between two neighbouring vertices is $\mu := R/\mathcal{N}$.
\\
Now, the classical geometry that we want to reproduce is described by a line element that, in these coordinates, reads
\begin{align}
ds^2=-N^2dt^2+a^2(dx^2+dy^2+dz^2)
\end{align}
with $N$ the lapse function and $a$ the scale factor encoding the spatial geometry. In Ashtekar-Barbero variables, we find for the connection and densitized triad respectively
\begin{align}
A^I_a = c \delta^I_a,\hspace{30pt} E^a_I=p \delta_I^a
\end{align}
with $c$ and $p$ being the fundamental variables. One can now compute the holonomy and the flux along each edge $e_I$ in direction $I$:
\begin{align}\label{coshol&flux}
h(e_I)= e^{-c \mu \tau_I}, \hspace{30pt} E^J(S_{e_I})= p \mu^2 n^J_{(I)}
\end{align}
where $\vec n_{(I)}$ is the unit vector normal to $S_{e_I}$ (so its components wrt cartesian coordinates are $n^J_{(I)} = \delta^J_I$). Therefore, according to (\ref{polar-decomposition}), the element $H_{I}\in \text{SL}(2,\mathbb{C})$ that should label the coherent state $\psi^t_{e_I, H_I} \in \mathcal H_{e_I}$ is (no sum over $I$)
\begin{align}
H_{I} & = \exp \left(-\frac{it}{\hbar\kappa\beta} p \mu^2 \tau_I\right) e^{-c \mu\tau_I} = \exp \left[\left(-2c \mu - i \frac{2t}{\hbar\kappa\beta} p \mu^2\right) \vec n_{(I)} \cdot \vec \tau/2\right] = \notag
\\
& = n_I \exp\left[\left(-2\mu c - i \frac{2\mu^2p}{a^2\beta}\right)\tau_3/2\right] n_I^{\dagger}
\end{align}
where in the third step we used the $SU(2)$-covariance of $\tau_I$ to move the rotation from its basis index to its matrix indices. In particular, $n_I$ are the $SU(2)$ elements that rotate the unit vector $\hat z$ into the unit vector $\vec n_{(I)}$, and are explicitly given by
\begin{align} \label{explicit-n}
n_1=\frac{1}{\sqrt{2}}\left(\begin{array}{cc}
1 & -1\\
1 & 1
\end{array}\right),\hspace{20pt}n_2=\frac{1}{\sqrt{2}}\left(\begin{array}{cc}
1 & i\\
i & 1
\end{array}\right),\hspace{20pt}n_3=\left(\begin{array}{cc}
1 & 0\\
0 & 1
\end{array}\right)
\end{align}
In this way, we have expressed $H_I$ in its {\it holomorphic decomposition}, which for a generic $SL(2,\mathbb C)$ element reads $n \exp(\bar{z} \tau_3/2) n'^{\dagger}$ for $z \in \mathbb C$ and $n,n' \in SU(2)$. While in general $z$, $n$ and $n'$ are independent, in this particularly simple case we find that $n = n'$ are fixed (though different for the three possible orientations of the edges) and we read off
\begin{align}\label{z-label}
z = -2\mu c + i \frac{2\mu^2p}{a^2\beta} \equiv \xi + i\eta
\end{align}
The complex number $z$ is therefore the only label of our coherent states, encoding the classical geometry described by the canonical pair $(c,p)$.
\\
Having the labels $\{h^{\mathbb C}\} = \{H\}$, we finally use (\ref{Appendix-formula for diagonal g}) to find our coherent states:
\begin{align}\label{CosCohSta}
\begin{array}{rl}
\Psi_{(c,p)}(\{g\}) & := \prod_{e\in E(\gamma)}\psi^t_{e,h_e^{\mathbb C}}(g_e) = \prod_{I\in\{1,2,3\}}\prod_{k\in\mathbb{Z}^3_{\mathcal{N}}}\psi_{I,H_I}(g_{k,I})
\\
\\
\psi_{I,H_I}(g) & := \dfrac{1}{\sqrt{\langle1\rangle_z}}\sum_{j\in\mathbb{N}/2}d_je^{-j(j+1)t/2}\sum_{m=-j}^je^{izm}D^{(j)}_{mm}(n_I^{\dagger}gn_I)
\end{array}
\end{align}
where $\langle 1 \rangle_z:=||\psi_{I,H_I}||^2$ is the normalization of the state and $\mathbb{Z}_{\mathcal N}=\{0,1...,\mathcal{N}-1\}$.
\\
These are states on the kinematical Hilbert space: we still have to implement the Gauss constraint and the vector constraint. We will not implement the Gauss constraint explicitly, as we are only interested in the expectation value of gauge-invariant observables, i.e., $\hat O_F$ such that $U_G[g]^{\dagger} \hat O_F U_G[g] = \hat O_F$ for any $SU(2)$ transformation $U_G[g]$. Moreover, it is easy to see by (\ref{gauge-variant coh}) that the coherent states are {\it gauge-covariant}: for a gauge transformation of $\tilde{g}$ at $x=s_e$, the starting point of edge $e$, we get $U_G[\tilde{g}]\psi^t_{e,h^{\mathbb{C}}_e}(g)=\psi^t_{e,h^{\mathbb{C}}_e}(g\tilde g)=\psi^t_{e,h^{\mathbb{C}}_e\tilde{g}^{\dagger}}(g)$. Combining both with the fact that the coherent states are sharply peaked, we get
\begin{align}
\langle \Psi^G_{(c,p)}, \hat{O}_F   \Psi^G_{(c,p)}\rangle &= \int D[\{\tilde{g}\}] \int D[\{\tilde{g}'\}]  \prod_{e\in E(\gamma)} \langle \psi_{e,H_e}, U[\tilde{g}_{s_e}]^{\dagger} U[\tilde{g}_{t_e}]^{\dagger} \hat{O}_F U[\tilde{g}'_{s_e}] U[\tilde{g}'_{t_e}] \psi_{e,H_e\tilde{g}'_{s_e}}\rangle =\nonumber\\
&= \int D[\{\tilde{g}\}] \int D[\{\tilde{g}'\}]  \prod_{e\in E(\gamma)} \langle \psi_{e,\tilde{g}_{t_e}H_e{\tilde{g}_{s_e}}^{-1}},  \hat{O}_F  \psi_{e,\tilde{g}'_{t_e}H_e{\tilde{g'}_{s_e}}^{-1}}\rangle\nonumber\\
&=\int D[\{\tilde{g}\}] \int D[\{\tilde{g}'\}]  \prod_{e\in E(\gamma)} \langle \psi_{e,\tilde{g}_{t_e}H_e{\tilde{g}_{s_e}}^{-1}},  \hat{O}_F  \psi_{e,\tilde{g}_{t_e}H_e{\tilde{g}_{s_e}}^{-1}}\rangle \delta(\tilde{g},\tilde{g}') + \mathcal{O}(t)=\nonumber\\
&=\int D[\{\tilde{g}\}]  \langle \Psi_{(c,p)}, U_G[\tilde{g}]^{\dagger} \hat{O}_F  U_G[\tilde{g}] \Psi_{(c,p)}\rangle +\mathcal{O}(t)= \nonumber\\
&=\langle \Psi_{(c,p)}, \hat{O}_F  \Psi_{(c,p)}\rangle +\mathcal{O}(t)
\end{align}
where in the last step we also used the fact that the Haar measure is normalized. This result guarantees that the expectation values have physical significance at leading order in $t$, without having to impose the Gauss constraint. More work is needed in order to incorporate the first order of quantum corrections (see appendix \ref{recovering-gauge} for a proof of principle).
\\
As for the vector constraint, the naive expectation is that working on an abstract graph automatically takes care of the diffeomorphisms. The situation, however, is more subtle. Indeed, in \cite{diffeo-merda} it was shown that states which appear to be orthogonal from the abstract graph perspective, actually are diffeomorphism-equivalent upon embedding, and therefore correspond to the same state in the diffeomoephism-invariant Hilbert space. This means that there remains a trace of the diffeomorphisms at the level of abstract lattices -- what could be called ``residual diffeomorphisms''. However, the argument presented in \cite{diffeo-merda} relies on the possibility that some links of the lattice are turned off (see fig. 3.1). On the other hand, the cosmological coherent states we just defined are ``maximal'', in the sense that they do not allow any link to be turned off: they live on the lattice $\mathbb Z_{\mathcal N}^3$ and {\it could not} live on any of its sublattices. This is due to the fact that the lattice is finite. If we considered infinite lattices, then there would be the possiblity of completely ``turning off'' one or more rows in it. However, the solution is simple: one simply needs to be aware that the state obtained by such a procedure is in fact identical to the original state. By this reasoning, we see that our cosmological coherent states are already invariant under the residual diffeomorphisms of the lattice.
\begin{figure}[H]
\begin{center}
\includegraphics{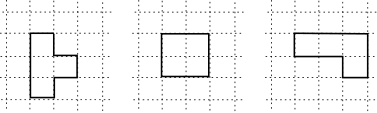}
\end{center}
\caption{\footnotesize Examples of graphs which are inequivalent on the abstract lattice, but are related by diffeormorphisms when embedded in the spatial manifold.}
\end{figure}
We therefore conclude that the states introduced above can be considered as physical states. The reminder of the section collects some general results about this particular subfamily, which will be used in the following sections to perform computations. Because the extension to many edges is trivial, we can focus on a single edge, and therefore we shall drop the index `$I$'. Moreover, we will sometimes write $H(z)$ to indicate that the $SL(2,\mathbb C)$ label $H$ effectively depends only on $z$ given in (\ref{z-label}).
		\subsection{General Properties of Cosmological Coherent States}
Consider $\psi_{e,H(z)}(g)$ as in (\ref{CosCohSta}), with $H(z)=n e^{z\tau_3/2}n^{\dagger}$ and $z=\xi+i\eta$ as in (\ref{z-label}). The first result gives us a way to simplify expectation values of operators involving left-invariant vector fields.
\\
{\bf Lemma 1:} Let $P(L,\hat{h})$ be a polynomial operator, with $L^K$ the left-invariant vector field. Then:
\begin{align}\label{Lemma1}
\langle \psi_{e,H(z)}, P(L(e),\hat{h}(e)) \psi_{e,H(z)}\rangle = \langle \psi_{e,H(-z)},P(-R(e),\hat{h}(e)^{\dagger}) \psi_{e,H(-z)}\rangle
\end{align}
\begin{proof}
Because of linearity, if suffices to consider a single basis element $\hat{h}_{a_1b_1}(e)^{r_1}L^{K_1}..\hat{h}_{a_nb_n}(e)^{r_n}L^{K_n}$ with $r_i\in\mathbb{N}_0$ and for arbitrary $j,j'$ in the defintion of $\psi_{e,H(z)}$. We recall that $\hat{h}$ is a multiplication operator, while for $R$ we find
\begin{align}\label{R-field-daggered}
R^K D^{(j)}_{ab}(g^{\dagger}) & =(-)^{b-a}R^KD^{(j)}_{-b-a}(g)=(-)^{b-a}\left.\frac{d}{ds}\right|_{s=0}D^{(j)}_{-b-a}(e^{s\tau_K}g) = \notag
\\
& = \left.\frac{d}{ds}\right|_{s=0}D^{(j)}_{ab}((e^{s\tau_K}g)^{\dagger}) = \left.\frac{d}{ds}\right|_{s=0}D^{(j)}_{ab}(g^\dag e^{-s\tau_K}) = \notag
\\
& = -\left.\frac{d}{ds}\right|_{s=0}D^{(j)}_{ab}(g^\dag e^{s\tau_K})
\end{align}
obtained using the properties of Wigner matrices, see appendix \ref{basic-properties}. In light of this, we have
\begin{align}
& \int d\mu_H(g) \overline{D^{(j')}_{m'n'}(n^{\dagger}gn)}D^{(\frac{1}{2})}_{a_1b_1}(g)^{r_1}L^{K_1}\ldots D^{(\frac{1}{2})}_{a_nb_n}(g)^{r_n}L^{K_n}D^{(j)}_{mn}(n^{\dagger}gn)\delta_{m'n'}\delta_{mn}e^{izm}e^{-i\bar{z}m'} = \notag
\\
& = \left.\frac{d}{ds_1}\right|_{s_1=0} ... \left.\frac{d}{ds_n}\right|_{s_n=0}\int d\mu_H(g) \overline{D^{(j')}_{m'm'}(n^{\dagger}gn)}D^{(\frac{1}{2})}_{a_1b_1}(g)^{r_1}D^{(\frac{1}{2})}_{a_2b_2}(ge^{s_1\tau_{K_1}})^{r_2} \times \notag
\\
&  \hspace{10pt}\times D^{(\frac{1}{2})}_{a_nb_n}(ge^{s_1\tau_{K_1}} ... e^{s_{n-1}\tau_{K_{n-1}}})^{r_n}D^{(j)}_{mm}(n^{\dagger}g^{\dagger}e^{s_1\tau_{K_n}}n)e^{izm}e^{-i\bar{z}m} = \notag
\\
& = (-1)^n \int d\mu_H(g^{\dagger}) \overline{D^{(j')}_{m'm'}(n^{\dagger}g^{\dagger}n)}D^{(\frac{1}{2})}_{a_1b_1}(g^{\dagger})^{r_1} R^{K_1} ... D^{(\frac{1}{2})}_{a_nb_n}(g^{\dagger})^{r_n}R^{K_n}D^{(j)}_{mm}(n^{\dagger}g^{\dagger}n)e^{izm}e^{-i\bar{z}m'} = \notag
\\
& = (-1)^n\int d\mu_H(g) \overline{D^{(j')}_{-m'-m'}(n^{\dagger}gn)}D^{(\frac{1}{2})}_{a_1b_1}(g^{\dagger})^{r_1} ... D^{(\frac{1}{2})}_{a_nb_n}(g^{\dagger})^{r_n}R^{K_n}D^{(j)}_{-m-m}(n^{\dagger}gn)e^{izm}e^{-i\bar{z}m'} = \notag
\\
& = (-1)^n \int d\mu_H(g)\overline{D^{(j')}_{m'm'}(n^{\dagger}gn)}D^{(\frac{1}{2})}_{a_1b_1}(g^{\dagger})R^{K_1} ... D_{a_nb_n}^{(\frac{1}{2})}(g^{\dagger})^{r_n}R^{K_n}D^{(j)}_{mm}(n^{\dagger}gn)e^{i(-z)m}e^{-i(-\bar{z})m'}
\end{align}
where in the second step we renamed the integration variable $g \rightarrow g^\dag$ and made use of (\ref{R-field-daggered}), in the third step we used $d\mu_H(g)=d\mu_H(g^{\dagger})$ and $D^{(j)}_{mn}(g^{\dagger})=\overline{D^{(j)}_{nm}(g)} = (-1)^{n-m} D^{(j)}_{-n-m}(g)$ (see appendix \ref{basic-properties}), and in the last we renamed $-m\rightarrow m, -m'\rightarrow m'$ (recall that sums over such indices are understood). This gives the statement.
\end{proof}
{\bf Lemma 2:} Let $M(R(e),\hat{h}(e))^{K_1,...,K_n}_{a_1b_1,...,a_{n'}b_{n'}}$ be a monomial operator, with index-structure stemming from $R^{K_i}(e)$ and $\hat{h}_{a_ib_i}(e)$ . Then:
\begin{align}\label{Lemma2}
& \langle \psi_{e,H(z)}, P(R(e),\hat{h}(e))^{K_1,...,K_n}_{a_1b_1,...,a_{n'}b_{n'}} \psi_{e,H(z)}\rangle =
D^{(1)}_{-K_1,-S_1}(n)\ldots D^{(1)}_{-K_n,-S_n}(n)\times\\
&\times D^{(\frac{1}{2})}_{a_1a_1'}(n)D^{(\frac{1}{2})}_{b_1'b_1}(n^{\dagger})\ldots D^{(\frac{1}{2})}_{a_{n'}a_{n'}'}(n)D^{(\frac{1}{2})}_{b_{n'}'b_{n'}}(n^{\dagger})
\langle \psi_{e,H(z)|_{n=1}}, P(R(e),\hat{h}(e))^{S_1,...,S_n}_{a_1'b_1',...,a_{n'}'b_{n'}'} \psi_{e,H(z)|_{n=1}}\rangle\nonumber
\end{align}
where we point out that $H(z)|_{n=1}=e^{z\tau_3/2}$.
\\
\begin{proof}
First, consider the action of $R^K$ on $D^{(j)}_{mn}(n^{\dagger}gn)$:
\begin{align} \label{RD-ngn}
R^K D^{(j)}_{mn}(n^{\dagger}gn) & = D^{(j)}_{mm'}(n^{\dagger}) \left(R^K D^{(j)}_{m'n'}(g)\right) D^{(j)}_{n'n}(n) = D^{(j)}_{mm'}(n^{\dagger}) D'^{(j)}_{m'\mu}(\tau_K) D^{(j)}_{\mu n'}(g) D^{(j)}_{n'n}(n) = \notag
\\
& = D^{(j)}_{mm'}(n^{\dagger}) D'^{(j)}_{m'\mu}(\tau_K) D^{(j)}_{\mu \nu}(n) D^{(j)}_{\nu n}(n^\dag gn) = \notag
\\
& = D^{(1)}_{-K-S}(n) D'^{(j)}_{m\nu}(\tau_S) D^{(j)}_{\nu n}(n^\dag gn)
\end{align}
where in the last step we used (\ref{tau-rotation}). Hence, for the generic monomial, we express numerous times the product of two holonomies as a linear combination with fixed coefficients $c$:
\begin{align}
& \int d\mu_H(g) \overline{D^{(j')}_{m'n'}(n^{\dagger}gn)}\hat{h}_{a_1b_1} ... R^{K_n} D^{(j)}_{mn}(n^{\dagger}gn) =
\\
& = D^{(\frac{1}{2})}_{a_1a_1'}(n)D^{(\frac{1}{2})}_{b_1'b_1}(n^{\dagger}) ... D^{(\frac{1}{2})}_{a_na_n'}(n)D^{(\frac{1}{2})}_{b'_nb_n}(n^{\dagger})\int d\mu_H(g)\overline{D^{(j')}_{m'n'}(n^{\dagger}gn)} \times \notag
\\
&\hspace{10pt} \times D^{(\frac{1}{2})}_{a_1'b_1'}(n^{\dagger}gn)R^{K_1} ... D^{(\frac{1}{2})}_{a_n'b_n'}(n^{\dagger}gn) \left( D^1_{-K_n-S_n}(n){D'}^j_{m\mu_n}(\tau_{S_n}) D^{(j)}_{\mu_nn}(n^{\dagger}gn)\right) = \notag
\\
& = D^{(\frac{1}{2})}_{a_1a_1'}(n)D^{(\frac{1}{2})}_{b_1'b_1}(n^{\dagger}) ... D^{(\frac{1}{2})}_{a_na_n'}(n)D^{(\frac{1}{2})}_{b'_nb_n}(n^{\dagger}) D^1_{-K_n-S_n}(n){D'}^j_{m\mu_n}(\tau_{S_n}) \times \notag
\\
& \hspace{10pt}\times \int d\mu_H(g) \overline{D^{(j')}_{m'n'}(n^{\dagger}gn)} D^{(\frac{1}{2})}_{a_1'b_1'}(n^{\dagger}gn)R^{K_1} ... R^{K_{n-1}}\sum_{j_n} c^n_{j_n,\mu_n'\nu_n}(\mu_n)D^{j_n}_{\mu_n'\nu_n'}(n^{\dagger}gn) \notag =
\\
& = D^{(\frac{1}{2})}_{a_1a_1'}(n)D^{(\frac{1}{2})}_{b_1'b_1}(n^{\dagger}) ... D^{(\frac{1}{2})}_{a_na_n'}(n)D^{(\frac{1}{2})}_{b'_nb_n}(n^{\dagger}) D^1_{-K_n-S_n}(n) ... D^1_{-K_1-S_1}(n) \times \notag
\\
& \hspace{10pt}\times \left({D'}^j_{m\mu_n}(\tau_{S_n}) ... {D'}^j_{m\mu_1}(\tau_{S_1}) \times \int d\mu_H(g) \overline{D^{(j')}_{m'n'}(g)}\sum_{j_n...j_1}
c^1_{j_1,\mu_1',\nu_1}(\mu_1)...c^n_{j_n,\mu_n'\nu_n}(\mu_n)D^{j_1}_{\mu_1'\nu_1'}(g)\right)\nonumber
\end{align}
where in the last line we used invariance of the Haar measure to replace $n^{\dagger}gn\rightarrow g$. We now see that the term in brackets is nothing but the expansion of $\int d\mu_H(g) \overline{D^{(j')}_{m'n'}(g)}\hat{h}_{a_1b_1} ... R^{K_n} D^{(j)}_{mn}(g)$, which was the statement.
\end{proof}
{\bf Lemma 3:} Let $P(R,L,\hat{h})$ be a polynomial operator on $\mathcal H_e$. Then:
\begin{align} \label{r&l-Switch}
\begin{array}{c}
\langle \psi_{e,H(z)|_{n=1}}, P(R,L,\hat{h}) L^K \psi_{e,H(z)|_{n=1}} \rangle = e^{-izK}
\langle \psi_{e,H(z)|_{n=1}}, P(R,L,\hat{h}) R^K \psi_{e,H(z)|_{n=1}} \rangle
\\
\\
\langle \psi_{e,H(z)|_{n=1}}, L^K P(R,L,\hat{h}) \psi_{e,H(z)|_{n=1}} \rangle = e^{-i\bar{z}K}
\langle \psi_{e,H(z)|_{n=1}}, R^K P(R,L,\hat{h}) \psi_{e,H(z)|_{n=1}} \rangle
\end{array}
\end{align}
\begin{proof}
Since ${D'}^j_{mn}(\tau_K)$ enforces $n+K-m=0$ (see appendix \ref{basic-properties}), one gets
\begin{align}
L^K D^{(j)}_{mm}e^{izm} & = D^{(j)}_{m\mu}(g){D'}^{(j)}_{\mu m}(\tau_K)e^{izm} = e^{-izK} D^{(j)}_{\mu m}(g){D'}^{(j)}_{m\mu}(\tau_K) e^{izm}= \notag
\\
& = e^{-izK} R^K D^{(j)}_{mm}(g)e^{izm}
\end{align}
where in the second step we exchanged the dummy indices $\mu \leftrightarrow m$. This is the first property. For the second, we expand $P(R,L,\hat{h})\psi=\sum_{j}c_{jmn}D^{(j)}_{mn}(g)$ with some coefficients $c$:
\begin{align}
& \sum_j c_{jmn} \int d\mu_H(g) \overline{D^{(j')}_{m'm'}(g)}e^{-i\bar{z}m'} L^K D^j_{mn}(g) = \frac{1}{d_{j'}} c_{j'm'n} e^{-i\bar{z}m'}{D'}^{j'}_{m'n}(\tau_K) = \notag
\\
& = e^{-i\bar{z}K}\frac{1}{d_{j'}}c_{j'nm'}e^{-i\bar{z}m'}{D'}^{(j')}_{nm'}(\tau_K) = \notag
\\
& = e^{-i\bar{z}K} \sum_j c_{jn\mu}e^{-i\bar{z}m'}{D'}^{j}_{n\nu}(\tau_K)\int d\mu_H(g)\overline{D^{(j')}_{m'm'}(g)}D^j_{\nu \mu}(g) = \notag
\\
& = e^{-i\bar{z}K} R^K P(R,L,\hat{h})\psi
\end{align}
having exchanged the dummy indices $m'\leftrightarrow n$ in the second step.
\end{proof}
Now, if one uses both relation in (\ref{r&l-Switch}), and the fact that $[R^K(e),L^M(e)]=0$, one gets immediately the following result.
\\
{\bf Corollary:} The following {\it cyclic property} holds:
\begin{align}\label{cyclicity}
\langle \psi_{e,H(z)|_{n=1}}, R^{K_1} .. R^{K_n} \psi_{e,H(z)|_{n=1}} \rangle = e^{-2\eta K_n}
\langle \psi_{e,H(z)|_{n=1}}, R^{K_n}R^{K_1} .. R^{K_{n-1}} \psi_{e,H(z)|_{n=1}} \rangle
\end{align}
where $z=\xi+i\eta$.
\\
As we will see in the next section, this property allows to greatly simplify the computations for the expectation value of any product os $R$'s.
\section{Expectation Values of Monomials on a Single Edge}
\label{s4}
In this section we will compute the expectation values of the various monomials which appear in the geometric operators. Thanks to lemma 2, it suffices to express everything on cosmological coherent states with $H(z)|_{n=1}$, and so we will use a shorthand notation for the {\it non-normalized} expectation values:
\begin{align}
\langle P(R(e),\hat{h}(e))\rangle_{z} & := \langle1\rangle_z \langle \psi_{e,H(z)|_{n=1}}, P(R(e),\hat{h}(e)) \psi_{e,H(z)|_{n=1}}\rangle
\end{align}
Moreover, we will change the basis of $\mathfrak{su}(2)$, meaning that instead of $I,J,K\in \{1,2,3\}$ we will consider the {\it spherical basis}, $s\in \{-,0,+\}$, where $\tau_{\pm}:=\mp (\tau_1 \pm i\tau_2)/\sqrt 2$ and $\tau_0:=\tau_3$. The generators are thus
\begin{align}
\tau_+ = i\sqrt{2}\left(\begin{array}{cc}
0 & 1\\
0 & 0
\end{array}\right),\hspace{20pt}
\tau_- = -i\sqrt{2}\left(\begin{array}{cc}
0 & 0\\
1 & 0
\end{array}\right)\hspace{20pt}\tau_0 = -i\left(\begin{array}{cc}
1 & 0\\
0 & -1
\end{array}\right)
\end{align}
subject to the algebra $[\tau_+,\tau_-]=2i\tau_0$, $[\tau_{\pm},\tau_0]=\pm 2i\tau_{\pm}$.\footnote
{
This does not change the action of geometric operators such as volume (\ref{ALvolume}), since they are by construction $SU(2)$-scalars, and hence invariant under any basis transformation.
}
		\subsection{Monomials of right-invariant Vector Field}
Consider first $N$ right-invariant vector fields, all with magnetic index $s_1=..=s_n=0$. We have
\begin{align}\label{MonRight}
& \langle R^{s_1} .. R^{s_N}\rangle_z = \notag
\\
& = \sum_{j,j'}d_jd_{j'}e^{-j(j+1)t/2}e^{-j'(j'+1)t/2}e^{-i\bar{z}m'}e^{izm}\int d\mu_H(g)\overline{D^{(j')}_{m'm'}(g)}{D'}^{(j)}_{m\mu_N}(\tau_0) .. {D'}^{(j)}_{\mu_2\mu_1}(\tau_0)D^{(j)}_{\mu_1m}(g)\notag
\\
& = \sum_{j,j'}d_jd_{j'}e^{-j(j+1)t/2-j'(j'+1)t/2}(-2im)^Ne^{-i\bar{z}m'}e^{izm}\int d\mu_H(g) \overline{D^{(j')}_{m'm'}(g)}D^{(j)}_{mm}(g) \notag
\\
& = \sum_j d_j e^{-j(j+1)t}(-2im)^Ne^{-2\eta m}=\sum_j d_j e^{-j(j+1)t}(i\partial_{\eta})^N e^{-2\eta m}=\nonumber\\
&=(i\partial_{\eta})^N\sum_j d_je^{-j(j+1)t}\frac{\sinh(d_j\eta)}{\sinh(\eta)}=(i\partial_{\eta})^N\langle 1\rangle_z
\end{align}
where we used ${D'}_{mn}(\tau_0)=-2im\delta_{mn}$ (see appendix \ref{basic-properties}) in the second step and the geometric sum $\sum_{m=-j}^j e^{-2\eta m}=\sinh(d_j\eta)/\sinh(\eta)$ to go to the last line. It remains to compute $\langle1\rangle_z$, the normalization of the state, for which we follow closely \cite{Winkler1,Winkler2,Winkler3}. As the authors there have pointed out, the elementary Poisson Summation Formula comes in handy.
\\
{\bf Theorem: (Poisson Summation Formula)} Consider $f\in L_1(\mathbb{R},dx)$  such that the series $\sum_{n\in\mathbb{Z}} f(y+ns)$ is absolutely and uniformly convergent for $y\in [0,s]$, $s>0$. Then
\begin{align}
\sum_{n\in\mathbb{Z}}f(ns)=\sum_{n\in\mathbb{Z}}\int_{\mathbb{R}}dx \cdot e^{-i2\pi nx}f(sx)
\end{align}
\begin{proof}
See e.g. the book about fourier analysis by Bochner \cite{Bochner}.
\end{proof}
By realizing that for $d_j=2j+1$ the term in the sum is even, we extend the sum to negative values, thus bringing $\langle1\rangle_z$ in the form to apply this theorem:
\begin{align}
\langle1\rangle_z & =\sum_{d_j=1}^{\infty} d_je^{-(d_j^2-1)t/4}\frac{\sinh(d_j\eta)}{\sinh(\eta)}=\frac{1}{2}\sum_{n=-\infty}^{\infty} n e^{-(n^2-1)t/4}\frac{\sinh(n\eta)}{\sinh(\eta)}=\notag
\\
& = \frac{1}{2}\int_{\mathbb{R}}du\sum_{n\in\mathbb{Z}}e^{-i2\pi n u}e^{-tu^2/4}e^{t/4}u\frac{\sinh(u\eta)}{\sinh(\eta)}
\end{align}
Upon completing the square, in the exponential one gets the term $e^{-4\pi^2n^2/t}$ which, for $t \rightarrow 0$, goes to $0$ faster than any polynomial, unless $n=0$. We conclude that, for $1\gg t$, only the $n=0$ term of the sum contributes, up to an error of order $\mathcal{O}(t^{\infty})$. We thus find
\begin{align}
\langle1\rangle_z=\frac{1}{2}e^{t/4}\int_{\mathbb{R}}du\ u e^{-tu^2/4}\frac{e^{2\eta u}}{\sinh(\eta)}=2e^{t/4}\sqrt{\frac{\pi}{t^3}}\frac{\eta e^{\eta^2/t}}{\sinh(\eta)}
\end{align}
Because of the factor $e^{\eta^2/t}$ in $\langle1\rangle_z$, the leading order of (\ref{MonRight}) in $t$ is obtained when all $N$ derivatives $\partial_{\eta}$ hit $e^{\eta^2/t}$, giving $\mathcal O(1/t^N)$.
\\
Let us now consider the case where some indices $s_1,...,s_n$ are not equal to zero. Since ${D'}^{(j)}_{\mu_{i+1}\mu_{i}}(\tau_{s_i})$ implies $\mu_{i+1}=\mu_i+s_i$ and we have $\mu_0=\mu_{N+1}=m$ it follows that $\sum_i s_i =0$. Consequently, a single non-vanishing $s_i$ is impossible: we shall therefore consider a pair $s_1, s_2$ with opposite sign. Moreover, we will neglect all contributions smaller than $\mathcal{O}(1/t^{N-1})$, since we saw that the leading order (for (\ref{MonRight})) is $\sim 1/t^N$. Using the algebra (for $s_1,s_2,s\neq 0$)
\begin{align}\label{SphAlgebra}
\left[ R^{s_1}, R^{s_2} \right] =-i (s_1-s_2) R^0,\hspace{20pt} \left[R^s,R^0\right]=-2isR^s
\end{align}
we find for the expectation value with a spacing $C$ between $s_1$ and $s_2$
\begin{align} \label{MonRspair}
\langle R^{0} ... R^{s_1} \stackrel{C}{\overbrace{R^0...R^0}}R^{s_2} ... R^{0}\rangle_z & = \langle R^{0} ... R^{0} R^{s_1} R^0...R^0R^{s_2}\rangle_z = \notag
\\
& = \langle \stackrel{N-2}{\overbrace{R^0...R^0}}R^{s_1}R^{s_2}\rangle_z-2iCs_2\langle \stackrel{N-3}{\overbrace{R^0...R^0}}R^{s_1}R^{s_2}\rangle_z+\mathcal{O}(1/t^{N-2}) = \notag
\\
& = \left((i\partial_{\eta})^{N-2}-2iCs_2(i\partial_{\eta})^{N-3}\right)\langle R^{s_1}R^{s_2}\rangle_z +\mathcal{O}(1/t^{N-2})
\end{align} 
having used (\ref{cyclicity}) in the first step and (\ref{SphAlgebra}) in the second. We reduced the problem to evaluating the expectation value $\langle R^{s_1}R^{s_2}\rangle_z$. But this can be done without effort by combining the cyclicity property and the algebra: it is
\begin{align}
\langle R^{s_1}R^{s_2}\rangle_z & = e^{-2\eta s_2}\langle R^{s_2}R^{s_1}\rangle_z=e^{-2\eta s_2}(\langle R^{s_1}R^{s_2}\rangle_z-\langle[R^{s_1},R^{s_2}]\rangle_z)=\nonumber
\\
&=e^{-2\eta s_2}\langle R^{s_1}R^{s_2}\rangle_z+e^{-2\eta s_2}i(s_1-s_2)\langle R^0\rangle_z=\nonumber
\\
&=e^{-2\eta s_2}\langle R^{s_1}R^{s_2}\rangle_z-e^{-2\eta s_2}(s_1-s_2)\partial_{\eta}\langle 1\rangle_z
\end{align}
which, solved for $\langle R^{s_1}R^{s_2}\rangle_z$, gives
\begin{align}
\langle R^{s_1}R^{s_2}\rangle_z=\frac{e^{-\eta s_2}}{\sinh(\eta)}\partial_{\eta}\langle1\rangle_z
\end{align}
Again, the leading order is obtained when all $\partial_{\eta}$ hit $e^{\eta^2/t}$. It follows that the term proportional to $C$ in (\ref{MonRspair}) is negligible, and the other is already next-to-leading wrt (\ref{MonRight}). Explicitly, we get
\begin{align}
\langle R^{0} ... R^{s_1} \stackrel{C}{\overbrace{R^0...R^0}}R^{s_2} ... R^{0}\rangle_z = -i \frac{e^{-\eta s_2}}{\sinh(\eta)}(i\partial_{\eta})^{N-1} \langle1\rangle_z +\mathcal{O}(1/t^{N-2})
\end{align} 
A similar calculation reveals that four and more non-vanishing indices are of order $\mathcal{O}(1/t^{N-2})$, and will thus be neglected.
\\
The final result up to linear quantum corrections thus read:
\begin{align}\label{ExpMonR}
& \langle R^{s_1}\ldots R^{s_N}\rangle_z =
\\
& = \left[\delta^{s_1...s_n}_0(i\partial_{\eta})^N-\frac{i}{\sinh(\eta)}\sum_{A<B=1}^N\delta^{s_1..\cancel s_A..\cancel s_B..s_N}_0 \left(\delta^{s_As_B}_{+1-1} e^{+\eta} + \delta^{s_As_B}_{-1+1} e^{-\eta}\right)(i\partial_{\eta})^{N-1}\right]\langle1\rangle_z \notag
\end{align}
Making use of lemma 1, equation (\ref{Lemma1}), one can straightforwardly generalize this result to a monomial in left-invariant vector fields:
\begin{align}\label{ExpMonL}
\langle L^{s_1}...L^{s_N}\rangle_z & = (-1)^N\langle R^{s_1}...R^{s_N}\rangle_{-z} = (-1)^{2N}\langle R^{s_N}...R^{s_1}\rangle_{z}=\notag
\\
& = \langle R^{s_N}...R^{s_1}\rangle_{z}
\end{align}
where in the second step we used the explicit expression (\ref{ExpMonR}) to find how a change in sign of $z$ (or $\eta$) influences the expectation value.
		\subsection{Monomials of Holonomy Operator}
As is well known from recoupling theory (see appendix \ref{basic-properties}), the product of Wigner matrices can be expressed as a linear combination of a single wigner matrix:
\begin{align}\label{recoupling}
D^{(j_1)}_{ab}(g)D^{(j_2)}_{cd}(g)=\sum_{j=|j_1-j_2|}^{j_1+j_2}d_j(-1)^{m-n}\left(\begin{array}{ccc}
j_1 & j_2 & j\\
a & c & m
\end{array}\right)\left(\begin{array}{ccc}
j_1 & j_2 & j\\
b & d & n
\end{array}\right) D^{(j)}_{-m-n}(g)
\end{align}
This property is extremely useful, since it allows to reduce the problem of computing $\langle \hat h_{a_1b_1} ... \hat h_{a_nb_n} \rangle_z$ to computing $\langle \hat h^{(j)}_{mn} \rangle_z$ (for the required values of $j$), by which we mean the operator whose action is to multiply by $D^{(j)}_{mn}(g)$.
\\
From the  explicit expression (\ref{CosCohSta}), we obtain (without normalization)
\begin{align}
\langle \hat h^{(k)}_{ab} \rangle_z & = \sum_{j,j'}d_jd_{j'}e^{-[j(j+1) + j'(j'+1)]t/2} e^{i(zm - \bar z m')} \int d\mu_H(g) \overline{D^{(j')}_{m'm'}(g)} D^{(k)}_{ab}(g) D^{(j)}_{mm}(g) = \notag
\\
& = \sum_{j,j'}d_jd_{j'}e^{-[j(j+1) + j'(j'+1)]t/2} e^{i\xi (m - m')} e^{-\eta(m+m')} \left(
\begin{array}{ccc}
j & k & j'
\\
m & a & -m'
\end{array}
\right) \left(
\begin{array}{ccc}
j & k & j'
\\
m & b & -m'
\end{array}
\right) = \notag
\\
& = \delta_{ab} e^{-i\xi a} \gamma^k_a
\end{align}
where in the second line we performed the integral (see (\ref{basic-recoupling-integral1}) and (\ref{basic-recoupling-integral2})), and in the third we used the observation that $a = m' - m = b$ to extract $e^{i\xi (m - m')} = e^{-i\xi a}$ from the sums and defined the quantity
\begin{align}
\gamma^k_{a}:=\sum_{j,j'} d_j d_{j'} e^{-t[j(j+1)+j'(j'+1)]/2}e^{-\eta(m+m')}\left(\begin{array}{ccc}
k & j & j'\\
a & m & -m'
\end{array}\right)^{2}
\end{align}
If we interchange in $\gamma^k_{a}$ the contracted indices $j\leftrightarrow j'$, $m\leftrightarrow m'$ everything is clearly invariant except for the $3j$-symbol:
\begin{align}
\left(\begin{array}{ccc}
k & j & j'\\
a & m & -m'
\end{array}\right)\rightarrow\left(\begin{array}{ccc}
k & j' &j\\
a & m' & -m
\end{array}\right)=\left(\begin{array}{ccc}
k & j & j'\\
-a & m & -m'
\end{array}\right)
\end{align}
As the index $a$ appeared only in the $3j$-symbol this leads to $\gamma^k_a\rightarrow\gamma^k_{-a}$, but since we only interchanged contracted indices $\gamma^k_a$ must stay invariant: we conclude that
\begin{align}
\gamma^k_a =\gamma^k_{-a}
\end{align}
The various values of $\gamma^k_a$ can now be computed with the Poisson Summation Formula. In appendix \ref{hol-int} the explicit computations are presented for $k=1/2$ and $k=1$ (which are relevant for the expectation value and dispersion of the holonomy operator, and sufficient for the Hamiltonian operator). The results are:
\begin{align} \label{gammas-result}
\begin{array}{rl}
\gamma^{1/2}_{1/2}&= \langle1\rangle_z \left[1+ \dfrac{t}{4\eta}\left(\dfrac{3}{4}\eta-\tanh\left(\dfrac{\eta}{2}\right)\right) + \mathcal O(t^2)\right]
\\
\\
\gamma^1_0 &=\langle 1 \rangle_z \left[1 + t \dfrac{2 \sinh(\eta/2)}{\eta \sinh(\eta)} + \mathcal O(t^2)\right]
\\
\\
\gamma^1_1 &=\langle 1 \rangle_z \left[1 - t \left(\dfrac{1}{4} + \dfrac{1}{2\eta} \tanh(\eta/2)\right) + \mathcal O(t^2)\right]
\end{array}
\end{align}
		\subsection{Holonomies and right-invariant Vector Fields}
In this section we present the strategy to compute expectation values of monomials involving both holonomy and right-invariant vector field. We consider a couple of explicit examples.
\\
Let us start with the commutator of an holonomy with $N$ right invariant vector fields. Using the algebra (\ref{OperatorAlgebra}) and dropping all terms of order $\mathcal{O}(1/t^{N-3})$ and lower (since the leading order is $\mathcal{O}(1/t^{N-1})$), we find
\begin{align}\label{ExpCommutator}
& \langle \hat{h}_{ac} [\hat{h}^{\dagger}_{cb}, R^{s_1} ... R^{s_N}] \rangle_z = \delta_{ac}\langle R^{s_1} ... R^{s_N}\rangle_z-\langle\hat{h}_{ab} R^{s_1} ... R^{s_N}\hat{h}^{\dagger}_{cb}\rangle_z = \notag
\\
& = \delta_{ab} \langle R^{s_1} ... R^{s_N}\rangle_z - \langle R^{s_1} \hat{h}_{ac} R^{s_2} ... R^{s_N} \hat{h}^{\dagger}_{cb}\rangle_z + D'^{(1/2)}_{ad}(\tau_{s_1}) \langle \hat{h}_{dc} R^{s_2} ... R^{s_N} \hat{h}^{\dagger}_{cb} \rangle_z = \notag
\\
& = \delta_{ab}\langle R^{s_1} ... R^{s_N}\rangle_z - \langle R^{s_1}R^{s_2} \hat{h}_{ac} ... R^{s_N} \hat{h}^{\dagger}_{cb}\rangle_z + D'^{(1/2)}_{ad}(\tau_{s_2}) \langle R^{s_1} \hat{h}_{dc} R^{s_3} ... R^{s_N} \hat{h}^{\dagger}_{cb} \rangle_z +\notag
\\
& + D'^{(1/2)}_{ad}(\tau_{s_1}) \langle R^{s_2} \hat{h}_{dc}R^{s_3} ... R^{s_N} \hat{h}^{\dagger}_{cb}\rangle_z - D'^{(1/2)}_{ae}(\tau_{s_1}) D'^{(1/2)}_{ed}(\tau_{s_2}) \langle\hat{h}_{dc} R^{s_3} ... \hat{h}^{\dagger}_{cb}\rangle_z = ... = \notag
\\
& = \sum_{A=1}^N D'^{(\frac{1}{2})}_{ab}(\tau_{s_A}) \langle R^{s_1} ... \cancel R^{s_A} ... R^{s_N}\rangle_z - \notag
\\
& - \sum_{A<B=1}^N D'^{(\frac{1}{2})}_{ac}(\tau_{s_A}) D'^{(\frac{1}{2})}_{cb}(\tau_{s_B}) \langle R^{s_1} ... \cancel R^{s_A} ... \cancel R^{s_B} ... R^{s_N}\rangle_z+\mathcal{O}(1/t^{N-3})
\end{align}
So such term can be brought back to expectation values of $R$'s only.
\\
The other type of mixed term is of the form $\hat{h}_{ab} R^{s_1} ... R^{s_N}$. From expression (\ref{CosCohSta}), we get (without normalization)
\begin{align}
\langle \hat{h}_{ab} R^{s_1} ... R^{s_N} \rangle_z & = e^{-i\bar{z}b} \sum_{j,j'} d_jd_{j'}e^{-t[j(j+1)+j'(j'+1)]/2} \times \\
& \times {D'}^{(j)}_{m\mu_N}(\tau_{s_N}) ... {D'}^{(j)}_{\mu_2\mu_1}(\tau_{s_1})\left(\begin{array}{ccc}
\frac{1}{2} & j & j'\\
a & \mu_1 & -m'
\end{array}\right)\left(\begin{array}{ccc}
\frac{1}{2} & j & j' \\
b & m & -m'
\end{array}\right)e^{-2\eta m}\nonumber
\end{align}
where we again used (\ref{recoupling}) and performed the group integral. As we did previously for monomials in $R$'s, let us consider the case $s_1 = ... = s_N = 0$ first. Using ${D'}^{(j)}_{mn}(\tau_0) =-2im\delta_{mn}$ (see appendix \ref{basic-properties}), it is easy to see that
\begin{align} \label{hR..R-allzero}
\langle\hat{h}_{ab}R^0\ldots R^0\rangle_z = e^{-\eta b}(i\partial_{\eta})^Ne^{nb}\langle \hat{h}_{ab}\rangle_z
\end{align}
which has leading order $\mathcal{O}(1/t^N)$. Next, we have the possibility of a single index being nonzero, as well as a pair. The order of these is next-to-leading wrt to (\ref{hR..R-allzero}). Indeed, using $[R^0, R^s]=2isR^s$ for $C\leq N$, we get
\begin{align}\label{holRsRs1}
\langle\hat{h}_{ab} \stackrel{C}{\overbrace{R^0...R^0}} R^s\stackrel{N-1-C}{\overbrace{R^0...R^0}}\rangle_z & = \langle\hat{h}_{ab}\stackrel{C-1}{\overbrace{R^0...R^0}}R^s\stackrel{N-C}{\overbrace{R^0...R^0}}\rangle_z+2i\langle\hat{h}_{ab}\stackrel{C-1}{\overbrace{R^0...R^0}}R^s\stackrel{N-1-C}{\overbrace{R^0...R^0}}\rangle_z = \notag
\\
& = \langle\hat{h}_{ab}R^s\stackrel{N-1}{\overbrace{R^0...R^0}}\rangle_z+\mathcal{O}(1/t^{N-2}) = \notag
\\
& = e^{-\eta b}(i\partial_{\eta})^{N-1}e^{\eta b}\langle \hat{h}_{ab}R^s\rangle_z+\mathcal{O}(1/t^{N-2})
\end{align}
and
\begin{align}\label{holRsRs2}
\langle\hat{h}_{ab}R^0...R^0&R^sR^0...R^0R^{s'}R^0...R^0\rangle_z=(i\partial_{\eta})^{N-2}\langle \hat{h}_{ab}R^sR^{s'}\rangle_z+\mathcal{O}(1/t^{N-2})
\end{align}
We thus reduced the problem to the evaluation of $\hat h_{ab} R^{s}$ and $\hat h_{ab} R^s R^{s'}$. Again, these can be computed by cleverly combining the cyclicity of lemma 3 with the algebra:
\begin{align}
\langle \hat{h}_{ab}R^s\rangle_z & = \langle R^s \hat{h}_{ab}\rangle_z-\langle [R^s,\hat{h}_{ab}]\rangle_z=e^{i\bar{z}s}\langle L^s \hat{h}_{ab}\rangle_z - D'^{(1/2)}_{ac}(\tau_s)\langle \hat{h}_{cb}\rangle_z = \notag
\\
& = e^{i\bar{z}s}\left(\langle \hat{h}_{ab}L^s\rangle_z+\langle [L^s,\hat{h}_{ab}]\rangle_z\right)- D'^{(1/2)}_{ac}(\tau_s) \langle \hat{h}_{cb}\rangle_z = \notag
\\
& = e^{i\bar{z}s}\left(^{-izs}\langle \hat{h}_{ab}R^s\rangle_z+D'^{(1/2)}_{cb}(\tau_s)\langle\hat{h}_{ac}\rangle_z\right)-D'^{(1/2)}_{ac}(\tau_s)\langle\hat{h}_{cb}\rangle_z = \notag
\\
& = e^{2\eta s}\langle \hat{h}_{ab} R^s\rangle_z+e^{i\bar{z}s/2}\left(e^{i\bar{z}s/2}D'^{(1/2)}_{cb}(\tau_s)\langle\hat{h}_{ac}\rangle_z-e^{-i\bar{z}s/2}D'^{(1/2)}_{ac}(\tau_s)\langle\hat{h}_{cb}\rangle_z\right)
\end{align}
leading to
\begin{align} \label{holRs}
\langle \hat{h}_{ab}R^s\rangle_z = \frac{se^{izs/2}}{2\sinh(\eta)}\left(e^{-i\bar{z}s/2}D'^{(1/2)}_{ac}(\tau_s)\langle\hat{h}_{cb}\rangle_z-e^{i\bar{z}s/2}\langle\hat{h}_{ac}\rangle_zD'^{(1/2)}_{cb}(\tau_s)\right)
\end{align}
A similar computation gives
\begin{align}
\langle \hat{h}_{ab} R^{s} R^{s'} \rangle_z & = -i \dfrac{e^{\eta s}}{\sinh(\eta)} \langle \hat{h}_{ab} R^0 \rangle_z + \notag
\\
& + \dfrac{s}{2\sinh(\eta)} e^{izs/2} \left(e^{-i\overline z s/2} D'^{(1/2)}_{ac}(\tau_s) \langle \hat{h}_{cb} R^{s'} \rangle_z - e^{i\overline z s/2} \langle \hat{h}_{ac} R^{s'} \rangle_z D'^{(1/2)}_{cb}(\tau_s)\right)
\end{align}
Now, since (\ref{holRsRs1}) involves only $N-1$ derivatives of $\eta$, we can only get an $\mathcal{O}(1/t^{N-1})$ contribution if all derivatives hit $e^{\eta^2/t}$ in the normalization appearing in $\langle\hat{h}_{ab}\rangle_z=\delta_{ab}e^{-i\xi a}\langle 1\rangle_z$ (which is correct at leading order).
\\
Using the same argument for (\ref{holRsRs2}), and putting the results together with the $s_1 = ... = s_N = 0$ case, we finally obtain
\begin{align}\label{ExpHolR}
\langle \hat{h}_{ab}R^{s_1}\ldots R^{s_n}\rangle_z & = \left[\delta^{s_1...s_N}_0 \delta_{a'b'} e^{-\eta b'} (i\partial_\eta)^N e^{\eta b'} \left(1 + \frac{t}{4\eta}\left(\frac{3}{4}\eta-\tanh\left(\frac{\eta}{2}\right)\right)\right)\right. - \notag
\\
& - \dfrac{\sinh(\eta/2)}{\sinh(\eta)} \left.\sum_{A = 1}^N \delta^{s_1...\cancel s_A ... s_N}_0 (\delta^{s_A}_{+1} + \delta^{s_A}_{-1}) e^{s_A\eta/2} D'^{(\frac{1}{2})}_{a'b'}(\tau^{s_A}) (i \partial_\eta)^{N-1}\right. - \notag
\\
& - i \dfrac{\delta_{a'b'}}{\sinh(\eta)} \left.\sum_{A<B = 1}^N \delta^{s_1...\cancel s_A ... \cancel s_B ... s_N}_0 (\delta^{s_As_B}_{+1-1} + \delta^{s_As_B}_{-1+1}) e^{s_A\eta} (i\partial_\eta)^{N-1}\right]  \langle 1\rangle_z
\end{align}
\section{Expectation Values of Geometric Operators}
\label{s5}
The tools developed in the previous section shall now be put into action. We start by computing the expectation value and dispersion of the fundamental operators, therefore discussing the physical interpretation of the semiclassicality parameter t. Afterwards, we investigate the geometric observable Volume.
		\subsection{Expectation Values and Spread of Holonomy- and Flux-Operators}
Of particular interest is the expectation value and dispersion of the holonomy operator. Using lemma 2 and the results of (\ref{gammas-result}), for the normalized expectation value of $\hat h_{ab} \equiv \hat h^{(\frac{1}{2})}_{ab}$ on endge $e$ oriented along $\vec n_{(I)}$ we find
\begin{align} \label{h-one-half}
\langle \psi_{e,H(z)}, \hat h_{ab} \psi_{e,H(z)} \rangle & = D^{(\frac{1}{2})}_{aa'}(n_I) D^{(\frac{1}{2})}_{b'b}(n_I^\dag) \frac{1}{\langle 1 \rangle_z} \langle \hat h^{(\frac{1}{2})}_{a'b'}\rangle_z = \notag
\\
& = \left[D^{(\frac{1}{2})}_{a\frac{1}{2}}(n_I) e^{-i\xi/2} D^{(\frac{1}{2})}_{\frac{1}{2}b}(n_I^\dag) + D^{(\frac{1}{2})}_{a-\frac{1}{2}}(n_I) e^{i\xi/2} D^{(\frac{1}{2})}_{-\frac{1}{2}b}(n^\dag_I)\right] \dfrac{1}{\langle 1 \rangle_z} \gamma^{1/2}_{1/2} = \notag
\\
& = D^{(\frac{1}{2})}_{ab}(n_I e^{\xi \tau_3/2} n_I^\dag)\left[1 + \frac{t}{4\eta}\left(\frac{3}{4}\eta-\tanh\left(\frac{\eta}{2}\right)\right) + \mathcal O(t^2)\right]
\end{align}
Recalling that $n_I e^{\xi \tau_3/2} n^\dag_I = e^{\xi \vec n_{(I)} \cdot \vec \tau/2} = e^{-\mu c \tau_I}$, we see that the leading order of this expectation value is exactly the classical holonomy along such an edge, $h(e_{I})$, when it is embedded in flat Robertson-Walker spacetime. For the dispersion, we have (no sum over $a, b$)
\begin{align}
& \langle \psi_{e,H(z)}, \hat h^{(1/2)}_{ab} \hat h^{(1/2)}_{ab} \psi_{e,H(z)} \rangle = d_0 \left(\begin{array}{ccc}
1/2 & 1/2 & 0 \\
a & a & -0
\end{array}\right)
\left(\begin{array}{ccc}
1/2 & 1/2 & 0\\
b & b & -0
\end{array}
\right)
\langle \psi_{e,H(z)}, \hat h^{(0)}_{00} \psi_{e,H(z)} \rangle + \notag
\\
& + d_1 (-1)^{n-m} \left(\begin{array}{ccc}
1/2 & 1/2 & 1 \\
a & a & -m
\end{array}\right)
\left(\begin{array}{ccc}
1/2 & 1/2 & 1\\
b & b & -n
\end{array}
\right)
\langle \psi_{e,H(z)}, \hat h^{(1)}_{mn} \psi_{e,H(z)} \rangle = \notag
\\
& = 3 \left(\begin{array}{ccc}
1/2 & 1/2 & 1 \\
a & a & -2a
\end{array}\right)
\left(\begin{array}{ccc}
1/2 & 1/2 & 1\\
b & b & -2b
\end{array}
\right)
\langle \psi_{e,H(z)}, \hat h^{(1)}_{2a,2b} \psi_{e,H(z)} \rangle
\end{align}
where we used the properties of $3j$-symbols to find that $m = 2a$ and $n = 2b$, and so $(-1)^{2(b-a)} = 1$ since $(b-a)$ is always integer. Using the explicit values of 3j-symbols, we obtain that $\langle (\hat h^{(1/2)}_{ab})^2 \rangle =
\langle \hat h^{(1)}_{2a,2b} \rangle$. This can again be computed from lemma 2 and (\ref{gammas-result}):
\begin{align} \label{h-one}
& \langle \psi_{e,H(z)}, \hat h^{1}_{mn} \psi_{e,H(z)} \rangle = \notag
\\
& = \dfrac{1}{\langle 1 \rangle_z} \left[D^{1}_{m0}(n_I) D^{1}_{0n}(n_I^\dag) \gamma^1_{0} + D^{1}_{m,+1}(n_I) D^{1}_{+1,n}(n_I^\dag) e^{-i\xi} \gamma^1_{1} + D^{1}_{m,-1}(n_I) D^{1}_{-1,n}(n_I^\dag) e^{i\xi} \gamma^1_{-1}\right] = \notag
\\
& = D^{1}_{mn}(n_I e^{\xi \tau_3/2} n_I^\dag) \dfrac{1}{\langle 1 \rangle_z} \gamma^1_{1} + D^{1}_{m0}(n_I) D^{1}_{0n}(n_I^\dag) \dfrac{1}{\langle 1 \rangle_z} (\gamma^1_{0} - \gamma^1_{1}) = \notag
\\
& = D^{1}_{mn}(n_I e^{\xi \tau_3/2} n_I^\dag) \left[1 - t \left(\frac{1}{4} + \frac{1}{2\eta} \tanh(\eta/2)\right)\right] + \notag
\\
& + n_{(I)}^m \overline{n_{(I)}^n} t \left[\frac{1}{4} + \frac{1}{2\eta} \left(\tanh(\eta/2) + \frac{4 \sinh(\eta/2)}{\sinh(\eta)}\right)\right] + \mathcal O(t^2)
\end{align}
where we used the fact that $D_{0m}^1(n_{I}) = n_{(I)}^m$, that is, the component $m$ of unit vector $\vec n_{(I)}$ in spherical basis. So the dispersion is finally
\begin{align}
\Delta h_{ab} & := \langle \psi_{e,H(z)}, (\hat h^{(1/2)}_{ab})^2 \psi_{e,H(z)} \rangle - \langle \psi_{e,H(z)}, \hat h^{(1/2)}_{ab} \psi_{e,H(z)} \rangle^2 =
\\
& = t \left\{-D^{1}_{2a,2b}(n_I e^{\xi\tau_3/2} n_I^\dag) \frac{5}{8} + n_{(I)}^{2a} \overline{n_{(I)}^{2b}} \left[\frac{1}{4} + \frac{1}{2\eta} \left(\tanh(\eta/2) + \frac{4 \sinh(\eta/2)}{\sinh(\eta)}\right)\right]\right\} \notag
\end{align}
The dispersion is linear in $t$, and so it goes to $0$ in the classical limit $t \rightarrow 0$.
\\
The other fundamental operator is the flux, that is proportional to the right-invariant vector field (see (\ref{fluxy})): from lemma 2 and (\ref{ExpMonR}), one immediately finds
\begin{align} \label{R-one}
\langle \psi_{e,H(z)}, R^{k} \psi_{e,H(z)} \rangle & = D^1_{-k-s}(n_I) \dfrac{1}{\langle 1 \rangle_z} \langle R^{s}\rangle_z = D^1_{-k-0}(n_I) \dfrac{1}{\langle 1 \rangle_z} i\partial_{\eta}\langle1\rangle = \notag
\\
& = \frac{2i\eta}{t} n_{(I)}^{-k} \left[1 + \frac{t}{2\eta^2} \left(1 - \eta \coth(\eta)\right)\right]
\end{align}
and similarly
\begin{align} \label{R-two}
& \langle \psi_{e,H(z)}, R^{k_1} R^{k_2} \psi_{e,H(z)} \rangle = D^1_{-k_1-s_1}(n_I) D^1_{-k_2-s_2}(n_I) \frac{1}{\langle 1 \rangle_z} \langle R^{s_1} R^{s_2}\rangle_z = \notag
\\
& = - \left(\frac{2\eta}{t}\right)^2 \left[n^{-k_1}_{(I)} n^{-k_2}_{(I)} \left[1 + \frac{t}{2\eta}\left(\frac{3}{\eta} - 2\coth\eta\right)\right]\right. - \notag
\\
& - \left. t \frac{1}{2\eta \sinh(\eta)}\left(D^{(1)}_{-k_1,-}(n_I) D^{(1)}_{-k_2,+}(n_I) e^{+\eta} + D^{(1)}_{-k_1,+}(n_I) D^{(1)}_{-k_2,-}(n_I) e^{-\eta}\right)\right]
\end{align}
where we used again the fact that $D^1_{m0}(n_I) = n_{(I)}^m$. At this point, we recall that these quantities are expressed in spherical basis. To recover the expectation values in cartesian basis, the following relations must be used:
\begin{align}
R^1 = \frac{(R^+ - R^-)}{\sqrt 2}, \ \ \ \ \ \ \ R^2 = -i\frac{(R^+ + R^-)}{\sqrt 2}, \ \ \ \ \ \ \ R^3 = R^0
\end{align}
Hence, one finds ($K \in \{1,2,3\}$)
\begin{align} \label{R-one-cartesian}
\langle \psi_{e,H(z)}, R^{K} \psi_{e,H(z)} \rangle = \dfrac{2i\eta}{t} n_{(I)}^K \left[1 + \dfrac{t}{2\eta^2} \left(1 - \eta \coth(\eta)\right)\right]
\end{align}
where $n^K_{(I)} = \delta^K_I$ are the cartesian components of $\vec n_{(I)}$ (we used the relation between spherical and cartesian components for vectors: $v^1 = (v^{-} - v^{+})/\sqrt{2}$, $v^2 = (-iv^{-} - iv^{+})/\sqrt{2}$ and $v^3 = v^0$). This equation shows that the cosmological state is peaked in the right-invariant vector field on the value $(2i\eta/t) \vec n_{(I)} = 2i\mu^2p/(\hbar \kappa \beta) \vec n_{(I)}$, which corresponds to the classical value of the flux: $E^J(S_{e_I}) = \mu^2 p \ n^J_{(I)}$. As for the dispersions, we first compute
\begin{align}
\langle \psi_{e,H(z)}, (R^{1})^2 \psi_{e,H(z)} \rangle & = \frac{1}{2} \langle \psi_{e,H(z)}, \left[(R^+)^2 + (R^-)^2 - R^+R^- - R^-R^+\right] \psi_{e,H(z)} \rangle = \notag
\\
& = -\left(\frac{2\eta}{t}\right)^2 \left[\left(\frac{n_{(I)}^- - n_{(I)}^+}{\sqrt 2}\right)^2 \left[1 + \frac{t}{2\eta}\left(\frac{3}{\eta} - 2\coth\eta\right)\right]\right. + \notag
\\
& + \left.t \frac{\coth\eta}{2\eta} \left(D^{(1)}_{++}(n_{I}) - D^{(1)}_{-+}(n_{I})\right) \left(D^{(1)}_{--}(n_{I}) - D^{(1)}_{+-}(n_{I})\right)\right] = \notag
\\
& = -\left(\frac{2\eta}{t}\right)^2 \left[\delta_{I}^1 \left[1 + \frac{t}{2\eta}\left(\frac{3}{\eta} - 2\coth\eta\right)\right] + t \frac{\coth\eta}{2\eta} \left(1 - \delta^1_I\right)\right] = \notag
\\
& = -\left(\frac{2\eta}{t}\right)^2 \left[\delta_{I}^1 + \delta_{I}^1 \frac{3t}{2\eta}\left(\frac{1}{\eta} - \coth\eta\right) + t \frac{\coth\eta}{2\eta}\right]
\end{align}
where in the second-to-last line we used the fact that $(D^{(1)}_{++}(n_{I}) - D^{(1)}_{-+}(n_{I}))(D^{(1)}_{--}(n_{I}) - D^{(1)}_{+-}(n_{I})) = |D^{(1)}_{++}(n_{I}) - D^{(1)}_{-+}(n_{I})|^2 = 1 - \delta^1_I$ (the last equality being cheched by explicit computation using (\ref{explicit-n})). A similar relation holds for $R^2$, while $R^3$ is immediately obtained from (\ref{R-two}). One then finds for the dispersions
\begin{align} \label{R-dispersions-cartesian}
\Delta R^K = \dfrac{2}{t} \left[n_{(I)}^K + (1 - n_{(I)}^K) \eta \coth(\eta)\right]
\end{align}
While it may be worrysome that this dispersion grows with $t \rightarrow 0$, this is expected since no quantum state can be infinitely peaked on both fundamental operators. What matters, however, is that the expectation value of $R^K$ also grows with $t \rightarrow 0$, and it does it in such a way that the ratio (i.e., the {\it relative dispersion}) actually tends to zero as $t \rightarrow 0$:
\begin{align}
\delta R^K := \left|\frac{\Delta R^K}{\langle \psi_{e,H(z)}, R^{K} \psi_{e,H(z)} \rangle^2}\right| = \dfrac{t}{2\eta^2} \left(1 + \dfrac{1 - n_{(I)}^K}{n_{(I)}^K} \eta \coth(\eta)\right)
\end{align}
		\subsection{Volume Operator}
We finally turn to the volume operator. Thanks to (\ref{Replacement}), the expectation value of Ashtekar-Lewandowski volume coincides with the expectation value of the ($k=1$)-Giesel-Thiemann volume operator (\ref{GTvolume}) up to next-to-leading order in $t$. But to evaluate that, we only need the expectation values of $\hat Q_v^N$ for $N = 1,2,4$ and $6$. Although these are operators on many edges, the expectation value reduces to the product of expectation values on each edge, so the only quantity we need is the expectation value of a string of $N$ right-invariant vector fields. This was derived in (\ref{ExpMonR}), and restoring the dependence on $n \in SU(2)$, it reads
\begin{align}\label{ExpRexpanded}
& \langle\psi_{e,H(z)}, R^{k_1} .. R^{k_N} \psi_{e,H(z)}\rangle = \left(\frac{2\eta i}{t}\right)^{N} D^{(1)}_{-k_1-s_1}(n) .. D^{(1)}_{-k_N-s_N}(n) \ (\delta_0^{s_1\ldots s_N} + 
\\
&\hspace{10pt} + \frac{t}{2\eta} [\delta_0^{s_1...s_N}\left(\frac{N(N+1)}{2\eta}-N\coth(\eta)\right)-\frac{1}{\sinh(\eta)}\sum_{A<B=1}^N\delta_0^{s_1..\cancel s_A ..\cancel s_B ...s_N}(\delta^{s_As_B}_{+1-1}+\delta^{s_As_B}_{-1+1})e^{s_A\eta}])\notag
\end{align}
In $\langle \hat{Q}_v^N \rangle$, one has a products of three such expectation values (one per every edge of the triple). The combinatorics is therefore encoded in $\epsilon_{k_ik'_ik''_i}R^{k_i}(e_1)R^{k_i'}(e_2)R^{k_i''}(e_3)$, which motivates us to consider the object
\begin{align}
\epsilon^{(n)}_{s_is'_is''_i} := \epsilon_{k_ik'_ik''_i} D^{(1)}_{-k_i-s_i}(n_1) D^{(1)}_{-k'_i-s'_i}(n_2) D^{(1)}_{-k'_i-s'_i}(n_3)
\end{align}
Since $n_i$ are fixed $SU(2)$ elements, the components of this tensor can be computed explicitly using (\ref{explicit-n}), and one in particular finds
\begin{align}
\epsilon^{(n)}_{00s} = \delta_{s0}
\end{align}
This is enough for our purposes: indeed, we are interested only in corrections linear in $t$, which means that two of the three strings in the product must be comprised only of $R$'s with vanishing index. $\epsilon^{(n)}_{00s}$ then forces the third index to also vanish, so one obtains
\begin{align}
\langle (R^0)^N \rangle_z = \delta_0^{s_1\ldots s_N} \left(\frac{2\eta i}{t}\right)^{N} \left[1 + \frac{t}{2\eta} \left(\frac{N(N+1)}{2\eta}-N\coth(\eta)\right)\right]
\end{align}
that is, only the terms proportional to $\delta^{s_1..s_N}_0$ will contribute.
\\
Now, the diffeomorphism-invariant quantity $\epsilon(e_a,e_b,e_c):=sgn(\det(a,b,c))=sgn(abc)\epsilon_{abc}$ with $a,b,c\in\{1,2,3\}$ tells us that (calling $R^I_a:=R^I(e_a)$)
\begin{align}
\langle\Psi_{(c,p)},\hat{Q}_v^N\Psi_{(c,p)}\rangle & = \langle\Psi_{(c,p)},	i^N\left(6\epsilon_{IJK} (R^I_1+R^I_{-1})(R^J_2+R^J_{-2})(R^K_3+R^K_{-3})\right)^N\Psi_{(c,p)}\rangle = \notag
\\
& = (6i)^N\prod_{i=1}^3\left(\sum_{n=0}^N \binom{N}{n} \langle (R^0_i)^n\rangle_z\langle (R^0_{-i})^{N-n}\rangle_z
\right) = \notag
\\
& = (6i)^N\left(\sum_{n=0}^N \binom{N}{n} \langle (R^0)^n \rangle_z\langle (R^0)^{N-n}\rangle_z
\right)^3 = \notag
\\
& = (6i)^N\left(\sum_{n=0}^N \binom{N}{n} \left(\frac{2\eta i}{t}\right)^N \left[1+\frac{t}{2\eta} \left(\frac{n(n+1)}{2\eta}-n\coth(\eta)\right)\right] \times \right. \notag
\\
&\hspace{10pt} \times \left.\left[1+\frac{t}{2\eta} \left(\frac{(N-n)(N-n+1)}{2\eta}-(N-n)\coth(\eta)\right)\right]\right)^3 = \notag
\\
& = (6i)^N\left(\frac{2\eta i}{t}\right)^{3N} \left[2^N+\frac{t}{2\eta^2}(N^2+3N)2^{N-2}-\frac{t}{2\eta}N2^N\coth(\eta)\right]^3
\end{align}
where we used
\begin{align}
\sum_{n=0}^N \binom{N}{n} = 2^N, \ \ \ \ \ \ \ \sum_{n=0}^N \binom{N}{n} n = 2^{N-1}N, \ \ \ \ \ \ \ \sum_{n=0}^N \binom{N}{n} n^2 = (N+N^2)2^{N-2}
\end{align}
Thus, we get
\begin{align}
\frac{\langle\Psi_{(c,p)},\hat{Q}_v^N\Psi_{(c,p)}\rangle}{\langle \Psi_{(c,p)},\hat{Q}_v\Psi_{(c,p)}\rangle^N} = 1+\frac{3t}{8\eta^2}N (N-1)
\end{align}
with which one can now compute the expectation value of the Giesel-Thiemann volume operator. For $k=1$, it reads
\begin{align}
\hat{V}^{GT}_{1,v} & = \frac{\langle\Psi_{(c,p)},\hat{Q}_v\Psi_{(c,p)}\rangle^{1/2}}{128} \times
\\
& \times \left[77\cdot\mathds{1}+77\frac{\hat{Q}_v}{\langle\Psi_{(c,p)},\hat{Q}_v\Psi_{(c,p)}\rangle^2}-33\frac{\hat{Q}_v^4}{\langle\Psi_{(c,p)},\hat{Q}_v\Psi_{(c,p)}\rangle^4}+7\frac{\hat{Q}^6_v}{\langle\Psi_{(c,p)},\hat{Q}_v\Psi_{(c,p)}\rangle^6}\right] \notag
\end{align}
so one finds (summing over all $\mathcal N^3$ vertices in the lattice)
\begin{align}
\langle \Psi_{(c,p)}, \hat{V}(\sigma) \Psi_{(c,p)} \rangle = \mathcal N^3 \sqrt{48} \left(\frac{2\eta}{t}\right)^{3/2} \left[1+\frac{3t}{4\eta^2} \left(\frac{7}{8} - \eta\coth(\eta)\right)+\mathcal{O}(t^2)\right]
\end{align}
\section{Conclusion}
\label{s6}
In this work we have constructed a family of coherent states in the full theory of LQC (on a cubic graph) based on gauge coherent states, and have shown that they are peaked on (discretized) flat Robertson-Walker cosmologies. These states are labelled by a parameter $\mu$ controlling how densely embedded is the graph in the spatial manifold. In order to approximate all observables which one needs to describe an isotropic universe, one should choose a sufficiently small $\mu$.
\\
We have presented all the necessary tools for computing the expectation values of any observable including corrections of first order in the semiclassicality parameter $t \sim \hbar$. This parameter is proportional to the spread of the coherent states and thus describes their quantum nature. In other words, this article provides the technology needed for the computation of observables including first order quantum corrections!
\\
We have shown that this works well for the example of the Ashtekar-Lewandowski volume, which can be recasted in polynomial form (instead of a square root) thanks to the result (\ref{Replacement}) by Giesel and Thiemann \cite{AQG3}. This replacement will also be central in the next paper of the series, where we shall turn our attention to the Hamiltonian operator. Specifically, we will use the tools presented here to compute the expectation value of the Hamiltonian on cosmological coherent states. In LQC it has been shown that, if one regards this expectation value as the effective Hamiltonian on the $(c,p)$-phase space, the corresponding effective dynamics agrees with the quantum evolution. Conjecturing that the same is true in LQG, it is important to evaluate this expectation value in the full theory and compare it with LQC. As already reported in \cite{DaporKlaus}, we will find that this expectation value does not coincide with the LQC effective Hamiltonian, due to the presence of the Lorentzian part in the Hamiltonian operator of the full theory.
\section*{Acknowledgements}
The authors would like to thank Kristina Giesel, Hans Liegener and Thomas Thiemann for helpful discussions. AD was partially supported by the Polish National Science Centre grant No. 2011/02/A/ST2/00300. KL thanks the German National Merit Foundation for financial support.
\appendix
\section{Expectation Value of Observables on gauge-invariant Coherent States}
\label{recovering-gauge}
We repeat the necessary definitions for gauge-invariance from main body of the article. There, we saw that one way to solve the Gauss constraint is by group averaging: this is a procedure which allows to obtain a gauge-invariant function $F_{\gamma}^G$ from any spin network function $F_{\gamma}$. Let $U_G[\{\tilde g\}]$ be the operator that generates a local transformation: a different $\tilde g \in SU(2)$ at every vertex of $\gamma$. Then,
\begin{align}
F^G_{\gamma}(\{g\})=\int D[\{\tilde{g}\}]U_G[\{\tilde{g}\}]F_{\gamma}(\{g\}) :=\left(\prod_{v\in V(\gamma)}\int d\mu_H(\tilde{g})\right) \prod_{e\in\gamma}f_e(\tilde{g}_{s_e}g\tilde{g}^{-1}_{t_e})
\end{align}
where $s_e$ (resp., $t_e$) denote the vertex at the beginning (resp., the end) of edge $e$.
\\
The coherent states $\psi^t_{e,H}$ on edge $e$ for $H \in SL(2,\mathbb{C})$ are not gauge-invariant, but {\it gauge-covariant}, which means
\begin{align} \label{covariant}
\psi^t_{e,H}(\tilde g_{s_e} g \tilde g_{t_e}^{-1}) = \psi^t_{e,\tilde{g}_{s_e}^{-1} H \tilde{g}_{t_e}}(g)
\end{align}
We shall now see what this says about expectation values of group-averaged coherent states.
\\
Let $M$ be a gauge-invariant monomial in holonomy operators and flux operators. It can be written as a product of monomials, one per each edge: $M = \prod_{e\in \gamma} M_e$, where $M_e$ only involves $h(e)$ and $E(e)$. Now, suppose we can compute the expectation value of $M_e$ on a coherent state peaked on $H \in SL(2,\mathbb{C})$:
\begin{align} \label{H-exp-H}
\langle \psi^t_{e, H}, M_e \psi^t_{e, H} \rangle = f^{(0)}(H) + t \ f^{(1)}(H) + \mathcal{O}(t^2)
\end{align}
where $f^{(0)}$ and $f^{(1)}$ are known functions. An important property of these analytic coherent states is that we can write the matrix elements of $M_e$ as
\begin{align} \label{H-exp-Hprime}
\langle \psi^t_{e, H}, M_e \psi^t_{e, H'} \rangle = f^{(0)}(H^{\mathbb{C}}) + t \ f^{(1)}(H^{\mathbb{C}}) + \mathcal{O}(t^2)
\end{align}
where $H^{\mathbb{C}} = H^{\mathbb{C}}(H,H')$ denotes the analytic continuation from $H$ to $H, H'$.
\\
On the other hand, from the sharp peakedness of coherent states, we know that there exists $\varphi_{H^{\mathbb C}}$ such that
\begin{align} \label{start}
\langle \psi^t_{e, H}, M_e \psi^t_{e, H'} \rangle & = \langle \psi^t_{e, H}, M_e \psi^t_{e, H} \rangle \delta(H,H') + t \langle \psi^t_{e, H}, M_e \varphi_{H^{\mathbb C}} \rangle + \mathcal{O}(t^2) = \notag
\\
& = f^{(0)}(H) \delta(H,H') + t \left[f^{(1)}(H) \delta(H,H') + \langle \psi^t_{e, H}, M_e \varphi_{H^{\mathbb C}} \rangle\right] + \mathcal{O}(t^2)
\end{align}
where we used (\ref{H-exp-H}) in the last step. Comparison with (\ref{H-exp-Hprime}) reveals that
\begin{align}
f^{(0)}(H^{\mathbb{C}}) = f^{(0)}(H) \delta(H,H'), \ \ \ \ \ \ \ \langle \psi^t_{e, H}, M_e \varphi_{H^{\mathbb C}} \rangle = f^{(1)}(H^{\mathbb{C}}) - f^{(1)}(H) \delta(H,H')
\end{align}
Now, inserting a resolution of identity on the lhs of the second equation, we find
\begin{align}
\langle \psi^t_{e, H}, M_e \varphi_{H^{\mathbb C}} \rangle & = \int d\nu_t(\tilde H) \langle \psi^t_{e, H}, M_e \psi^t_{e, \tilde H} \rangle \langle \psi^t_{e, \tilde H}, \varphi_{H^{\mathbb C}} \rangle = \notag
\\
& = \int d\nu_t(\tilde H) \langle \psi^t_{e, H}, M_e \psi^t_{e, H} \rangle \delta(H,\tilde H) \langle \psi^t_{e, H}, \varphi_{H^{\mathbb C}}\rangle + \mathcal O(t) = \notag
\\
& = f^{(0)}(H) \langle \psi^t_{e, H}, \varphi_{H^{\mathbb C}}\rangle + \mathcal O(t)
\end{align}
which leads to
\begin{align} \label{varphi-expr}
\langle \psi^t_{e, H}, \varphi_{H^{\mathbb C}}\rangle = \dfrac{f^{(1)}(H^{\mathbb{C}}) - f^{(1)}(H) \delta(H,H')}{f^{(0)}(H)} + \mathcal O(t)
\end{align}
Now, let us plug this back in the first line of (\ref{start}):
\begin{align}
\langle \psi^t_{e, H}, M_e \psi^t_{e, H'} \rangle & = \langle \psi^t_{e, H}, M_e \psi^t_{e, H} \rangle \delta(H,H') + t \langle \psi^t_{e, H}, M_e \varphi_{H^{\mathbb C}} \rangle + \mathcal{O}(t^2) = \notag
\\
& = \langle \psi^t_{e, H}, M_e \psi^t_{e, H} \rangle \delta(H,H') + t f^{(0)}(H) \langle \psi^t_{e, H}, \varphi_{H^{\mathbb C}} \rangle + \mathcal{O}(t^2) = \notag
\\
& = \langle \psi^t_{e, H}, M_e \psi^t_{e, H} \rangle \left[\delta(H,H') + t \langle \psi^t_{e, H}, \varphi_{H^{\mathbb C}}\rangle\right] + \mathcal{O}(t^2) = \notag
\\
& = \langle \psi^t_{e, H}, M_e \psi^t_{e, H} \rangle \left[\delta(H,H') + t \dfrac{f^{(1)}(H^{\mathbb{C}}) - f^{(1)}(H) \delta(H,H')}{f^{(0)}(H)}\right] + \mathcal{O}(t^2)
\end{align}
where in the third line we used that $f^{(0)}(H) = \langle \psi^t_{e, H}, M_e \psi^t_{e, H} \rangle$ at leading order. Having the matrix elements of $M_e$, we can finally compute the expectation value on group-averaged coherent states:
\begin{align}
\frac{\langle \Psi^G_{\gamma, \{H\}}, M \Psi^G_{\gamma, \{H\}}\rangle}{||\Psi^G_{\gamma, \{H\}}||^2} = \frac{\int_{SU(2)}D[\{\tilde{g}\}]\int_{SU(2)}D[\{\tilde{g}'\}]\langle \Psi^t_{\gamma, \{H\}}, U_G[\{\tilde{g}\}]^{\dagger} M U_G[\{\tilde{g}'\}] \Psi^t_{\gamma, \{H\}}\rangle}{\int_{SU(2)}D[\{\tilde{g}\}]\int_{SU(2)}D[\{\tilde{g}'\}]\langle \Psi^t_{\gamma,\{H\}}, U_G[\{\tilde{g}\}]^{\dagger} U_G[\{\tilde{g}'\}]\Psi^t_{\gamma,\{H\}}\rangle}
\end{align}
The denominator, which we call $Vol[\{H\}]$, can be computed once and for all (as it is independent of operator $M$). As for the numerator, using the covariance property (\ref{covariant}) we get
\begin{align}
\int_{SU(2)}D[\{\tilde{g}\}] & \int_{SU(2)}D[\{\tilde{g}'\}]\langle \Psi^t_{\gamma, \{H\}}, U_G[\{\tilde{g}\}]^{\dagger} M U_G[\{\tilde{g}'\}] \Psi^t_{\gamma, \{H\}} \rangle = \notag
\\
& = \int_{SU(2)} D[\{\tilde g\}] \int_{SU(2)} D[\{\tilde g'\}]\prod_{e\in \gamma} \langle \psi^t_{e,\tilde g_{s_e}^\dag H \tilde g_{t_e}}, M_e \psi^t_{e,\tilde g'^\dag_{s_e} H \tilde g'_{t_e}} \rangle =
\\
& = \int_{SU(2)}D[\{\tilde{g}\}] \prod_{e\in \gamma} \langle \psi^t_{e,\tilde g_{s_e}^\dag H \tilde g_{t_e}}, M_e \psi^t_{e,\tilde g^\dag_{s_e} H \tilde g_{t_e}} \rangle \times \notag
\\
& \times \left(1 + t \int_{SU(2)}D[\{\tilde{g}'\}]\frac{f^{(1)}(H^{\mathbb{C}})-f^{(1)}(\tilde g_{s_e}^\dag H \tilde g_{t_e})}{f^{(0)}(\tilde g_{s_e}^\dag H \tilde g_{t_e})} +\mathcal{O}(t^2)\right)
\end{align}
where $H^{\mathbb{C}} = H^{\mathbb{C}}(\tilde g_{s_e}^\dag H \tilde g_{t_e},\tilde g'^\dag_{s_e} H \tilde g'_{t_e})$. Now, we use covariance in the other direction and the fact that $M$ is gauge-invariant, to write
\begin{align}
\prod_{e} \langle \psi^t_{e,\tilde g^\dag_{s_e} H \tilde g_{t_e}}, M_e \psi^t_{e,\tilde g^\dag_{s_e} H \tilde g_{t_e}} \rangle & = \prod_e \langle \psi^t_{e,H}, U_G[\{\tilde{g}\}]^\dag M_e U_G[\{\tilde{g}\}]\psi^t_{e,H} \rangle = \notag
\\
& = \langle \Psi^t_{\gamma,\{H\}}, U_G[\{\tilde{g}\}]^\dag M U_G[\{\tilde{g}\}] \Psi^t_{\gamma,\{H\}} \rangle = \notag
\\
& = \langle \Psi^t_{\gamma,\{H\}}, M \Psi^t_{\gamma,\{H\}} \rangle
\end{align}
So we finally find for the expectation value on gauge-invariant coherent states
\begin{align}
\frac{\langle \Psi^G_{\gamma, \{H\}}, M \Psi^G_{\gamma, \{H\}}\rangle}{||\Psi^G_{\gamma, \{H\}}||^2} & = \frac{1}{Vol[\{H\}]} \langle \Psi^t_{\gamma,\{H\}}, M \Psi^t_{\gamma,\{H\}} \rangle \times
\\
& \times \left(1 + t \int_{SU(2)}D[\{\tilde{g}\}]\int_{SU(2)}D[\{\tilde{g}'\}] \frac{f^{(1)}(H^{\mathbb{C}})-f^{(1)}(\tilde g_{s_e}^\dag H \tilde g_{t_e})}{f^{(0)}(\tilde g^\dag_{s_e} H \tilde g_{t_e})} + \mathcal{O}(t^2)\right) \notag
\end{align}
Note that, to explicitly compute this, one must perform the integrals over many copies of $SU(2)$, which depend explicitly on the form of $M$ and thus require knowledge of its matrix elements on the basis of (non-gauge-invariant) coherent states.

\section{Properties of Wigner Matrices, right- and left-invariant Vector Fields, and $3j$-symbols}
\label{basic-properties}
In this appendix we shall present some general properties of the objects we have been using (most results and their proofs can be found in \cite{Brink-Satchler}). For the sake of clarity, we here restore the sums over magnetic indices.
\\
Let us start with the Wigner matrices, $D^{(j)}_{mn}$. The most important properties that they satisfy are that, for every $g \in SU(2)$,
\begin{align}
D^{(j)}_{mn}(g g') = \sum_\mu D^{(j)}_{m\mu}(g) D^{(j)}_{\mu n}(g'), \ \ \ \ \ D^{(j)}_{mn}(g^\dag) = [D^{(j)}_{mn}(g)]^\dag \equiv \overline{D^{(j)}_{nm}(g)}
\end{align}
with $j \in \{0,1/2,1,...\}$ and $m,n \in \{-j,-j+1,...,j\}$. These characterize $D^{(j)}$ as a matrix representation of $SU(2)$. For an $SU(2)$ element
\begin{align}
g = \left(\begin{array}{cc} a & b \\ c & d \end{array}\right)
\end{align}
the explicit formula for $D^{(j)}$ found to be (see \cite{Thi07} chapter 32)
\begin{align} \label{WignerD-explicit}
D^{(j)}_{mn}(g) = \sum_l \dfrac{\sqrt{(j+m)!(j-m)!(j+n)!(j-n)!}}{(j+n-l)!(m-n+l)!l!(j-m-l)!} a^{j+n-l} b^{m-n+l} c^l d^{j-m-l}
\end{align}
where the sum is over all $l$'s such that the arguments of the factorials are defined. This formula can be straightforwardly extended to the complexification of $g$, i.e., elements $H \in SL(2,\mathbb C)$. Then, in particular, for a diagonal element, $H = \exp(z\tau_3/2) = \text{diag}(e^{-iz/2},e^{+iz/2})$, (\ref{WignerD-explicit}) greatly simplifies:
\begin{align} \label{Appendix-formula for diagonal g}
D_{mn}(e^{z\tau_3/2}) = \delta_{mn}e^{-izm}
\end{align}
This formula is used in the main text to obtain the explicit form of cosmological coherent states: there, one has $\text{Tr}^{(j)}(H^\dag g)$ with $H = n e^{\bar z\tau_3/2} n^\dag$. Using cyclicity of the trace, one rewrites this as $\text{Tr}^{(j)}(e^{-z\tau_3/2} n^\dag g n) = D^{(j)}_{mn}(e^{-z\tau_3/2}) D^{(j)}_{nm}(n^\dag g n) = e^{izm} D^{(j)}_{mm}(n^\dag g n)$, where we used (\ref{Appendix-formula for diagonal g}) in the last step.
\\
Other properties used in the main text are
\begin{itemize}
\item hermitian conjugation:
\begin{align}
\overline{D^{(j)}_{mn}(g)} = (-1)^{m-n} D^{(j)}_{-m-n}(g)
\end{align}
\item orthogonality:
\begin{align} \label{D-matrices-orthogonality}
\int d\mu_H(g) \overline{D^{(j)}_{mn}(g)} D^{(j')}_{m'n'}(g) = \dfrac{1}{d_j} \delta^{jj'} \delta_{mm'} \delta_{nn'}
\end{align}
\item completeness (Peter-Weyl Theorem):
\begin{align}
f(g) = \sum_{j,m,n} c_{jmn} D^{(j)}_{mn}(g), \ \ \ \ \ c_{jmn} := d_j \langle D^{(j)}_{mn}, f \rangle = d_j \int d\mu_H(g) \overline{D^{(j)}_{mn}(g)} f(g)
\end{align}
for all $f \in L_2(SU(2), d\mu_H)$
\end{itemize}
Wigner matrices are also related to $3j$-symbols. Indeed, the latter can be defined as the coefficients of an $SU(2)$-invariant $3$-tensor: as such, they satisfy
\begin{align} \label{3j-rotation-invariance}
\left(
\begin{array}{ccc}
j & j' & j''
\\
m & m' & m''
\end{array}
\right) = \sum_{n,n',n''} D_{mn}^{(j)}(g) D_{m'n'}^{(j')}(g) D_{m''n''}^{(j'')}(g) \left(
\begin{array}{ccc}
j & j' & j''
\\
n & n' & n''
\end{array}
\right)
\end{align}
for all $g \in SU(2)$. $3j$-symbols have a number of important properties:
\begin{itemize}
\item selection rule (or triangular inequality):
\begin{align}
\left(
\begin{array}{ccc}
j & j' & J
\\
m & m' & M
\end{array}
\right) = 0 \ \text{unless} \ |j-j'| \leq J \leq j+j' \ \text{and} \ m + m' + M = 0
\end{align}
and of course $m \in \{j, j-1, ..., -j\}$, $m' \in \{j', j'-1, ..., -j'\}$, and $M \in \{J, J-1, ..., -J\}$
\item orthogonality:
\begin{align}
\begin{array}{l}
\sum_{J,M} d_J \left(
\begin{array}{ccc}
j & j' & J
\\
m & m' & M
\end{array}
\right) \left(
\begin{array}{ccc}
j & j' & J
\\
n & n' & M
\end{array}
\right) = \delta_{mn} \delta_{m'n'}
\\
\\
\sum_{m,m'} \left(
\begin{array}{ccc}
j & j' & J
\\
m & m' & M
\end{array}
\right) \left(
\begin{array}{ccc}
j & j' & J'
\\
m & m' & M'
\end{array}
\right) = \dfrac{1}{d_J} \delta^{JJ'} \delta_{MM'}
\end{array}
\end{align}
\item even permutation of columns (or cyclicity):
\begin{align}
\left(
\begin{array}{ccc}
j & j' & J
\\
m & m' & M
\end{array}
\right) = \left(
\begin{array}{ccc}
j' & J & j
\\
m' & M & m
\end{array}
\right) = \left(
\begin{array}{ccc}
J & j & j'
\\
M & m & m'
\end{array}
\right)
\end{align}
\item odd permutation of columns:
\begin{align}
\left(
\begin{array}{ccc}
j & j' & J
\\
m & m' & M
\end{array}
\right) = (-1)^{j+j'+J} \left(
\begin{array}{ccc}
j' & j & J
\\
m' & m & M
\end{array}
\right)
\end{align}
\item sign-swap:
\begin{align}
\left(
\begin{array}{ccc}
j & j' & J
\\
m & m' & M
\end{array}
\right) = (-1)^{j+j'+J} \left(
\begin{array}{ccc}
j & j' & J
\\
-m & -m' & -M
\end{array}
\right)
\end{align}
\end{itemize}
Combining (\ref{3j-rotation-invariance}) with the properties of Wigner matrices, we find
\begin{align} \label{DDJ-to-JDbar}
\sum_{n,n'} D_{mn}^{(j)}(g) D_{m'n'}^{(j')}(g) \left(
\begin{array}{ccc}
j & j' & J
\\
n & n' & N
\end{array}
\right) = \sum_M \left(
\begin{array}{ccc}
j & j' & J
\\
m & m' & M
\end{array}
\right) \overline{D^{(J)}_{MN}(g)}
\end{align}
Contracting now both sides with an appropriate $3j$-symbol, we can make use of the orthogonality property to obtain the following useful relation:
\begin{align}
D_{mn}^{(j)}(g) D_{m'n'}^{(j')}(g) = \sum_{J,M,N} d_J \left(
\begin{array}{ccc}
j & j' & J
\\
m & m' & M
\end{array}
\right) \left(
\begin{array}{ccc}
j & j' & J
\\
n & n' & N
\end{array}
\right) \overline{D^{(J)}_{MN}(g)}
\end{align}
This formula is at the heart of recoupling theory, and it is fundamental for our analysis, since it allows to rewrite a product of holonomies in terms of a single Wigner matrix (of higher spin), which then in turn enables us to use (\ref{D-matrices-orthogonality}). The first example of this in the main text is in the evaluation of the expectation value of a single holonomy (of spin $k$), $\hat h^{(k)}_{ab}$. In that case, one deals with an integral of the form
\begin{align} \label{basic-recoupling-integral1}
I_{mam',nbn'}^{(jkj')} := \int d\mu_H(g) \overline{D^{(j')}_{m'n'}(g)} D^{(k)}_{ab}(g) D^{(j)}_{mn}(g)
\end{align}
Now, one simply recouples the second and third matrices, and uses the hermitian conjugation property for the resulting matrix:
\begin{align} \label{basic-recoupling-integral2}
I_{mam',nbn'}^{(jkj')} & = \sum_{J,M,N} d_J (-1)^{M-N} \left(
\begin{array}{ccc}
k & j & J
\\
a & m & M
\end{array}
\right) \left(
\begin{array}{ccc}
k & j & J
\\
b & n & N
\end{array}
\right) \int \overline{D^{(j')}_{m'n'}(g)} D^{(J)}_{-M-N}(g) = \notag
\\
& = (-1)^{n'-m'} \left(
\begin{array}{ccc}
j & k & j'
\\
m & a & -m'
\end{array}
\right) \left(
\begin{array}{ccc}
j & k & j'
\\
n & b & -n'
\end{array}
\right)
\end{align}
where in the second step we performed the integral, consumed the various Kronecker delta's, and performed two odd permutations of columns (which gives $(-1)^{2(k+j+j')} = 1$, since the triangular inequality ensures that $k + j + j'$ is integer).
\\
Until now we provided basic properties and tools about Wigner matrices. The other set of objects which appears in the main text are right-invariant (and left-invariant) vector fields, $R^K$. In (\ref{R-and-D-prime}) we wrote the action of $R^K$ on a Wigner matrix as
\begin{align} \label{R-action-to-be-proven}
R^KD^{(j)}_{mn}(g) = \sum_\mu {D'}^{(j)}_{m\mu}(\tau_K)D^{(j)}_{\mu n}(g)
\end{align}
and declared that
\begin{align} \label{D-prime-definition}
{D'}^{(j)}_{mn}(\tau_K) = 2i\sqrt{j(j+1)d_j}(-1)^{j+n}\left(\begin{array}{ccc}
j & 1 & j\\
n & K & -m
\end{array}\right)
\end{align}
First of all, we must point out that, while the symbol $D'$ can be made sense of in any basis, equation (\ref{D-prime-definition}) holds only in the spherical one (since the $3j$-symbol on the right imposes $K \in \{0, \pm 1\}$). We shall therefore prove that (\ref{D-prime-definition}) is indeed correct in the spherical basis.
\begin{proof}
First, recall that $R^Kf(g)=(d/ds)_{s=0} f(e^{s\tau_K}g)$, so replacing $f$ with $D^{(j)}_{mn}$, we get
\begin{align}
R^KD^{(j)}_{mn}(g) & = \left(\frac{d}{ds}\right)_{s=0} D^{(j)}_{mn}(e^{s\tau_K}g) = \sum_{\mu} \left(\frac{d}{ds}\right)_{s=0} D^{(j)}_{m\mu}(e^{s\tau_K}) D^{(j)}_{\mu n}(g)
\end{align}
so comparison with (\ref{R-action-to-be-proven}) reveals that
\begin{align} \label{D-prime-as-derivative}
D'^{(j)}_{m\mu}(\tau_K) = \left(\frac{d}{ds}\right)_{s=0} D^{(j)}_{m\mu}(e^{s\tau_K})
\end{align}
so we are left with the task of performing this derivative. Given $\tau_{\pm}$ and $\tau_0$, it is easy to see that $e^{s\tau_K} = \sum_{n = 0}^\infty s^n (\tau_K)^n/n!$ gives
\begin{align}
e^{s\tau_{+}} = \left(\begin{array}{cc} 1 & is\sqrt 2 \\ 0 & 1 \end{array}\right), \ \ \ \ \ \ \ \ \ \ e^{s\tau_{-}} = \left(\begin{array}{cc} 1 & 0 \\ -is\sqrt 2 & 1 \end{array}\right), \ \ \ \ \ \ \ \ \ \ e^{s\tau_0} = \left(\begin{array}{cc} e^{-is} & 0 \\ 0 & e^{is} \end{array}\right)
\end{align}
Consider for instance $\tau_{+}$. From (\ref{D-prime-as-derivative}) and (\ref{WignerD-explicit}), we have
\begin{equation}
D'^{(j)}_{mn}(\tau_+) = \sum_l \dfrac{\sqrt{(j+m)!(j-m)!(j+n)!(j-n)!}}{(j+n-l)!(m-n+l)!l!(j-m-l)!} \left.\frac{d}{ds}\right|_{s=0} b^{m-n+l} c^l
\end{equation}
with $b = is\sqrt 2$ -- so that the derivative in $s$ gives $i\sqrt 2 (m-n+l) b^{m-n+l-1}$ -- and $c = 0$. The only way for this not to vanish is to have $l = 0$ (so that the sum collapses in a single term) and $m-n-1=0$, i.e., $\delta_{m,n+1}$. For these values, the factorials simplify, and the final formula is
\begin{equation}
D'^{(j)}_{mn}(\tau_+) = i\sqrt 2 \sqrt{(j+n+1)(j-n)} \delta_{m,n+1}
\end{equation}
Similar computations reveal that
\begin{equation}
D'^{(j)}_{mn}(\tau_-) = -i\sqrt 2 \sqrt{(j-n+1)(j+n)} \delta_{m,n-1}
\end{equation}
and
\begin{equation}
D'^{(j)}_{mn}(\tau_0) = -2in \delta_{mn}
\end{equation}
It is convenient to write these in terms of $3j$-symbols, as in this way we have a unique formula and the symmetry properties of these objects are apparent. Recalling the explicit formulae for $3j$-symbols
\begin{align}
\begin{array}{c}
\left(\begin{array}{ccc} j & 1 & j \\ n & 0 & -m \end{array}\right) = \delta_{mn} (-1)^{j+n+1} \dfrac{n}{\sqrt{j(j+1)d_j}}
\\
\\
\left(\begin{array}{ccc} j & 1 & j \\ n & \pm 1 & -m \end{array}\right) = \pm \delta_{m,n \pm 1} (-1)^{j+n} \sqrt{\dfrac{(j \mp n)(j \pm n + 1)}{2j(j+1)d_j}}
\end{array}
\end{align}
one can therefore write
\begin{equation}
D'^{(j)}_{mn}(\tau_K) = 2i \sqrt{j(j+1)d_j} (-1)^{j + n} \left(\begin{array}{ccc}
j & 1 & j\\
n & K & -m
\end{array}\right)
\end{equation}
which is exactly (\ref{D-prime-definition}).
\end{proof}
It is interesting to notice that, setting $j = 1/2$, one finds
\begin{align}
D'^{(1/2)}_{ab}(\tau_+) = \left(
\begin{array}{cc}
0 & 0 \\
-i\sqrt 2 & 0
\end{array}
\right), \ \ \ D'^{(1/2)}_{ab}(\tau_-) = \left(
\begin{array}{cc}
0 & i\sqrt 2 \\
0 & 0
\end{array}
\right), \ \ \ D'^{(1/2)}_{ab}(\tau_0) = i \left(
\begin{array}{cc}
1 & 0 \\
0 & -1
\end{array}
\right)
\end{align}
which are not quite $\tau_K$ themselves. The relation is
\begin{align}
D'^{(1/2)}(\tau_K) = \epsilon^\dag \tau_K \epsilon, \ \ \ \ \ \ \ \ \ \ \epsilon := \left(
\begin{array}{cc}
0 & 1 \\
-1 & 0
\end{array}
\right)
\end{align}
In light of (\ref{D-prime-definition}), it is easy to see that the following property holds
\begin{align} \label{tau-rotation}
\sum_{m',n'} D_{mm'}^{(j)}(g^\dag) D'^{(j)}_{m'n'}(\tau_K) D_{n'n}^{(j)}(g) = \sum_L D^{(1)}_{-K-L}(g) D'^{(j)}_{mn}(\tau_L)
\end{align}
as it descends from property (\ref{DDJ-to-JDbar}) of $3j$-symbols.
\section{Holonomy Integrals}
\label{hol-int}
In this appendix we want to explicitly compute
\begin{align}
\gamma^k_{a} := \sum_{j,j'\geq 0} d_j d_{j'} e^{-t [j(j+1)+j'(j'+1)]/2}e^{-\eta(m+m')}\left(\begin{array}{ccc}
k & j & j'\\
a & m & -m'
\end{array}\right)^2
\end{align}
for $\gamma^{1/2}_{1/2}$, $\gamma^1_0$ and $\gamma^1_1$ in particular. For this computation, we will require the following 3j-smybols:
\begin{align}\label{3jtable}
\begin{array}{lr}
\left(
\begin{array}{ccc}
j & j+\frac{1}{2} & \frac{1}{2}\\
m & -m-\frac{1}{2} & \frac{1}{2}
\end{array}
\right)^2=\dfrac{j + m + 1}{d_{j + 1/2}d_j}, &
\\
\\
\left(
\begin{array}{ccc}
j & j & 1\\
m & -m & 0
\end{array}
\right)^2=\dfrac{m^2}{j(j+1)d_j}, &
\left(
\begin{array}{ccc}
j & j& 1\\
m & -m-1 & 1
\end{array}
\right)^2=\dfrac{(j-m)(j+m+1)}{2j(j+1)d_j}
\\
\\
\left(
\begin{array}{ccc}
j & j+1 & 1\\
m & -m & 0
\end{array}
\right)^2=\dfrac{(j-m+1)(j+m+1)}{(j+1)d_jd_{j+1}}, &
\left(
\begin{array}{ccc}
j & j+1 & 1\\
m & -m-1 & 1
\end{array}
\right)^2=\dfrac{(j+m+1)(j+m+2)}{d_jd_{j+1/2}d_{j+1}}
\end{array}
\end{align}
Moreover, we will also need in addition to the standard Gaussian integrals (which can be done in the usual way by completing the square), the following non-trivial ones (for $a, b \sim t$):
\begin{align}\label{Int-sh/u}
\int^{\infty}_{-\infty} du \ e^{-au^2}\frac{\sinh(u)}{u} = \pi \text{Erfi}\left(\frac{1}{2\sqrt{a}}\right)
\end{align}
and
\begin{align}\label{Int-sh/u+a}
\int^{\infty}_{-\infty} du \ e^{-au^2+bu}\frac{\sinh(u)}{u} &
= \pi \ \text{Erfi}\left(\frac{1}{2\sqrt{a}}\right) + e^{\frac{1}{4a}} 2 \sqrt{\pi a} \left(\cosh\left(\frac{b}{2a}\right) - 1\right) + \mathcal O(t)
\end{align}
Let us show how the latter is derived.
\begin{proof}
First, write $e^{-au^2+bu} \sinh(u) = (e^{-au^2+bu+u}-e^{-au^2+bu-u})/2$ and change the integration variable in the second term as $u \rightarrow -u$, so that we obtain
\begin{align}
\int^{\infty}_{-\infty} du \ e^{-au^2+bu}\frac{\sinh(u)}{u}=\int^{\infty}_{-\infty} du \ e^{-au^2+u}\frac{\cosh(bu)}{u}
\end{align}
Now, expanding $\cosh$ in series, we have
\begin{align}
& \int^{\infty}_{-\infty} du \ e^{-au^2+bu}\frac{\sinh(u)}{u} =\int^{\infty}_{-\infty} du \ e^{-au^2+u}\frac{1}{u} + \sum_{n = 1}^\infty \dfrac{b^{2n}}{(2n)!} \int^{\infty}_{-\infty} du \ e^{-au^2+u} u^{2n-1} = \notag
\\
& =\int^{\infty}_{-\infty} du \ e^{-au^2}\frac{\sinh(u)}{u} + e^{\frac{1}{4a}} \sum_{n = 1}^\infty \dfrac{b^{2n}}{(2n)!} \int^{\infty}_{-\infty} du \ e^{-a(u-\frac{1}{2a})^2} u^{2n-1} = \notag
\\
& =\pi \ \text{Erfi}\left(\frac{1}{2\sqrt{a}}\right) + \dfrac{e^{\frac{1}{4a}}}{\sqrt a} \sum_{n = 1}^\infty \dfrac{b^{2n}}{(2n)!} \left(\frac{1}{2a}\right)^{2n-1} \int^{\infty}_{-\infty} dx \ e^{-x^2} \left(2 \sqrt a x + 1\right)^{2n-1}
\end{align}
where in the last step we have changed the integration variable in the second integral to $x = \sqrt{a} (u-1/(2a))$. Now, since we are interested in an expansion in powers of $a$ (as $a$ will be proportional to the small parameter $t$), we expand $\left(2 \sqrt a x + 1\right)^{2n-1} \approx 1 + (2n-1) 2\sqrt a x + (2n-1)(2n-2) 2ax^2$, and use standard Gaussian integrals to find
\begin{align}
& \sum_{n = 1}^\infty \dfrac{b^{2n}}{(2n)!} \left(\frac{1}{2a}\right)^{2n-1} \int^{\infty}_{-\infty} dx \ e^{-x^2} \left(2 \sqrt a x + 1\right)^{2n-1} = \notag
\\
& = b \sum_{n = 1}^\infty \dfrac{1}{(2n)!} \left(\frac{b}{2a}\right)^{2n-1} \sqrt \pi [1 + (2n-1)(2n-2)a] = \notag
\\
& = b \sqrt\pi \left[\frac{2a}{b} (\cosh(b/2a) - 1) + \frac{b}{2} \cosh(b/2a) - 2a \sinh(b/2a) + \frac{4a^2}{b} (\cosh(b/2a) - 1)\right] = \notag
\\
& = b \sqrt\pi \left[\frac{2a}{b} (\cosh(b/2a) - 1) + \mathcal O(a,b)\right]
\end{align}
where in the last step we used the fact that $a, b \sim t$, so the first term is the dominant one. Putting together, we then find
\begin{align}
\int^{\infty}_{-\infty} du \ e^{-au^2+bu}\frac{\sinh(u)}{u} &
= \pi \ \text{Erfi}\left(\frac{1}{2\sqrt{a}}\right) + e^{\frac{1}{4a}} 2 \sqrt{\pi a} \left(\cosh\left(\frac{b}{2a}\right) - 1\right) + \mathcal O(t)
\end{align}
which is the claim.
\end{proof}
We can now compute $\gamma^{1/2}_{1/2}$, $\gamma^1_0$ and $\gamma^1_1$. We start from $\gamma^1_0$. It is a sum of three terms: $j' = j$, $j' = j + 1$ and $j' = j - 1$ (the last one defined only if $j \geq 1$).
\begin{align}
\gamma^1_0 & = \sum_{j\geq 0}\frac{d_j}{j(j+1)}e^{-t(j+1)j}e^{-2\eta m} m^2+\notag
\\
& + \sum_{j\geq 0} e^{-t[j(j+1) + (j+1)(j+2)]/2}e^{-2\eta m} \frac{(j-m+1)(j+m+1)}{(j+1)}+\notag
\\
& + \sum_{j\geq 1} e^{-t[j(j+1)+j(j-1)]/2}e^{-2\eta m}\frac{j^2-m^2}{j}
\end{align}
After replacing the dummy index $j$ with $J = j - 1$ in the last term, it reduces to the second term. So, using $\sum_m e^{-2\eta m} m^\alpha = (-1/2 \partial_\eta)^\alpha \sinh(d_j \eta)/\sinh(\eta)$ and $u = d_j$, we get
\begin{align}
\gamma^1_0 & = \sum_{u\geq 1}\frac{u}{u^2-1} e^{-t(u^2-1)/4} \partial_\eta^2 \frac{\sinh(\eta u)}{\sinh(\eta)}+\notag
\\
& + \sum_{u\geq 1} e^{-t(j+1)^2} \frac{1}{u+1} ((u+1)^2 - \partial_\eta^2) \frac{\sinh(\eta u)}{\sinh(\eta)}
\end{align}
Now, both sums can be extended to $u \leq -1$ (which produces an extra $1/2$ factor), and the $u = 0$ terms can be added since it vanishes anyway. Hence, we can use Poisson formula to turn the sums into integrals:
\begin{align}
\gamma^1_0 & = \dfrac{1}{2} \int_{-\infty}^\infty du \ \frac{u}{u^2-1} e^{-t(u^2-1)/4} \partial_\eta^2 \frac{\sinh(\eta u)}{\sinh(\eta)} + \notag
\\
& + \dfrac{1}{2} \int_{-\infty}^\infty du \ e^{-t(u+1)^2/4} \frac{1}{u+1} ((u+1)^2 - \partial_\eta^2) \frac{\sinh(\eta u)}{\sinh(\eta)}
\end{align}
Consider the first integral. Taking one of the derivatives, we can rewrite it as
\begin{align}
& \dfrac{1}{2} \partial_\eta \left[\int_{-\infty}^\infty du \ \frac{u}{u^2-1} e^{-t(u^2-1)/4} \partial_\eta \frac{\sinh(\eta u)}{\sinh(\eta)}\right] = \notag
\\
& = \dfrac{1}{2} \partial_\eta \left[\int_{-\infty}^\infty du \ e^{-t(u^2-1)/4} \frac{u}{\sinh(\eta)^2} \frac{u \sinh(\eta) \cosh(u \eta) - \sinh(u\eta) \cosh(\eta)}{(u-1)(u+1)}\right] = \notag
\\
& = \dfrac{1}{2} \partial_\eta \left[\int_{-\infty}^\infty du \ e^{-t(u^2-1)/4} \frac{u}{2\sinh(\eta)^2} \left(\frac{\sinh((u+1)\eta)}{u + 1} - \frac{\sinh((u-1)\eta)}{u - 1}\right)\right] = \notag
\\
& = \partial_\eta \left[\int_{-\infty}^\infty du \ e^{-t(u^2-1)/4} \frac{u}{2\sinh(\eta)^2} \frac{\sinh((u+1)\eta)}{u + 1}\right] = \notag
\\
& = \partial_\eta \left[\frac{1}{2\sinh(\eta)^2} \int_{-\infty}^\infty dv \ e^{-t(v^2 -2v)/4} \frac{v - 1}{v} \sinh(v\eta)\right]
\end{align}
where we renamed $u \rightarrow -u$ in the second term of the second-to-last line (so it coincided with the first), and in the last line we chagned integration variable to $v = u + 1$. The integral is now reduced to a sum of two terms, both of which are of the forms presented above:
\begin{align}
& \dfrac{1}{2} \partial_\eta \left[\int_{-\infty}^\infty du \ \frac{u}{u^2-1} e^{-t(u^2-1)/4} \partial_\eta \frac{\sinh(\eta u)}{\sinh(\eta)}\right] = \notag
\\
& = \frac{1}{2} \partial_\eta \left[\frac{1}{\sinh(\eta)^2} \int_{-\infty}^\infty dv \ e^{-t(v^2 -2v)/4} \sinh(v\eta) - \frac{1}{\sinh(\eta)^2} \int_{-\infty}^\infty dv \ e^{-t(v^2 -2v)/4} \frac{1}{v} \sinh(v\eta)\right] = \notag
\\
& = \frac{1}{2} \partial_\eta \left[2\sqrt{\frac{\pi}{t}} e^{t/4} \frac{e^{\eta^2/t}}{\sinh(\eta)} - \dfrac{\pi}{\sinh(\eta)^2 } \text{Erfi}(\eta/\sqrt t) - e^{\eta^2/t} \dfrac{\sqrt{\pi t}}{\eta \sinh(\eta)^2} (\cosh(\eta) - 1)\right] = \notag
\\
& = 2 \sqrt{\frac{\pi}{t^3}} \frac{\eta e^{\eta^2/t}}{\sinh(\eta)} \left[1 + t \left(\frac{1}{4} - \frac{\coth(\eta)}{\eta}\right) + \mathcal O(t^2)\right]
\end{align}
where in the last step we performed the derivative and expanded the result in $t$, retaining only up to next-to-leading order. As for the second integral in $\gamma^1_0$, upon changing variable to $v = u + 1$ we get
\begin{align}
& \dfrac{1}{2} \int_{-\infty}^\infty dv \ e^{-tv^2/4} \frac{1}{v} (v^2 - \partial_\eta^2) \frac{\sinh(\eta (v - 1))}{\sinh(\eta)} = \notag
\\
& = \int_{-\infty}^\infty dv \ e^{-tv^2/4} \frac{1}{v} \frac{1}{\sinh(\eta)^2} (v \cosh(v\eta) - \coth(\eta) \sinh(v\eta)) = \notag
\\
& = \frac{1}{\sinh(\eta)^2} \left[2 e^{\eta^2/t} \sqrt{\frac{\pi}{t}} - \coth(\eta) \pi \text{Erfi}(\eta/\sqrt t)\right] = \notag
\\
& = 2 \sqrt{\frac{\pi}{t^3}} \frac{\eta e^{\eta^2/t}}{\sinh(\eta)} \left[t \frac{1}{\eta \sinh(\eta)} + \mathcal O(t^2)\right]
\end{align}
where again we retained only up to next-to-leading order in $t$. We can then write
\begin{align}
\gamma^1_0 & = 2 \sqrt{\frac{\pi}{t^3}} \frac{\eta e^{\eta^2/t}}{\sinh(\eta)} \left[1 + t \left(\frac{1}{4} - \frac{\coth(\eta)}{\eta} + \frac{1}{\eta \sinh(\eta)}\right) + \mathcal O(t^2)\right] = \notag
\\
& = 2 \sqrt{\frac{\pi}{t^3}} \frac{\eta e^{\eta^2/t}}{\sinh(\eta)} \left[1 + t \left(\frac{1}{4} + \frac{2 \sinh(\eta/2)}{\eta \sinh(\eta)}\right) + \mathcal O(t^2)\right]
\end{align}
\\
Let us now consider $\gamma^1_1$. Here we also have three terms:
\begin{align}
\gamma^1_1 & = \sum_{j\geq 0}\frac{d_j}{2j(j+1)} e^{-t(j+1)j}e^{-\eta (2m+1)} (j-m)(j+m+1) +\notag
\\
& + \sum_{j\geq 0} \dfrac{1}{2(j+1)} e^{-t[j(j+1) + (j+1)(j+2)]/2} e^{-\eta (2m+1)} (j+m+1)(j+m+2)+\notag
\\
& + \sum_{j\geq 1} \dfrac{1}{2j} e^{-t[j(j+1)+j(j-1)]/2}e^{-\eta (2m+1)} (j-m)(j-m-1)
\end{align}
The first one, written in terms of $u$, reads
\begin{align}
& \frac{e^{-\eta}}{2} \sum_{u\geq 1}\frac{u}{u^2-1} e^{-t(u^2-1)/4} \left[(u^2 - 1) + (2 - \partial_\eta)\partial_\eta\right] \frac{\sinh(u\eta)}{\sinh(\eta)} = \notag
\\
& = \frac{e^{-\eta}}{4} \int_{-\infty}^\infty du \ u e^{-t(u^2-1)/4} \frac{\sinh(u\eta)}{\sinh(\eta)} + \frac{e^{-\eta}}{4} (2 - \partial_\eta) \int_{-\infty}^\infty du \ \frac{u}{u^2-1} e^{-t(u^2-1)/4} \partial_\eta \frac{\sinh(u\eta)}{\sinh(\eta)} = \notag
\\
& = \sqrt{\frac{\pi}{t^3}} \frac{\eta e^{\eta^2/t}}{\sinh(\eta)} e^{t/4} e^{-\eta} + \notag
\\
& \hspace{10pt}+ \frac{e^{-\eta}}{4} (2 - \partial_\eta) \left[2\sqrt{\frac{\pi}{t}} e^{t/4} \frac{e^{\eta^2/t}}{\sinh(\eta)} - \dfrac{\pi}{\sinh(\eta)^2 } \text{Erfi}(\eta/\sqrt t) - e^{\eta^2/t} \dfrac{\sqrt{\pi t}}{\eta \sinh(\eta)^2} (\cosh(\eta) - 1)\right] = \notag
\\
& = 2\sqrt{\frac{\pi}{t^3}} \frac{\eta e^{\eta^2/t}}{\sinh(\eta)} e^{-\eta} \left[\frac{t}{2\eta} (1 + \coth(\eta)) + \mathcal O(t^2)\right]
\end{align}
where in the second-to-last step we used for the second integral the fact that it is exactly the same as before. Now, the second and third terms in $\gamma^1_1$ can be combined as one:
\begin{align}
& e^{-\eta} \sum_{u\geq 1} \left[\dfrac{e^{-t(u+1)^2/4}}{u+1} \left(\frac{u+1}{2} + m\right) \left(\frac{u+3}{2} + m\right) + \frac{e^{-t(u-1)^2/4}}{u-1} \left(\frac{u-1}{2} - m\right) \left(\frac{u-3}{2} - m\right)\right] e^{-2\eta m} = \notag
\\
& = \frac{e^{-\eta}}{4} \sum_{u=-\infty}^\infty \left[\dfrac{e^{-t(u+1)^2/4}}{u+1} \left(u+1 - \partial_\eta\right) \left(u+3 - \partial_\eta\right)\right] \frac{\sinh(\eta u)}{\sinh(\eta)} = \notag
\\
& = \frac{e^{-\eta}}{4} \int_{-\infty}^\infty dv \dfrac{e^{-tv^2/4}}{v} \left(v - \partial_\eta\right) \left(v + 2 - \partial_\eta\right) \frac{\sinh(\eta (v - 1))}{\sinh(\eta)} = \notag
\\
& = \frac{e^{-\eta}}{4} \int_{-\infty}^\infty dv \ e^{-tv^2/4} \left(v + 2 - \partial_\eta\right) \frac{\sinh(\eta (v - 1))}{\sinh(\eta)} - \notag
\\
& \hspace{10pt}- \frac{e^{-\eta}}{4} \int_{-\infty}^\infty dv \dfrac{e^{-tv^2/4}}{v} \left(v + 2 - \partial_\eta\right) \frac{1}{\sinh(\eta)^2} \left[\sinh(\eta) \cosh(\eta(v-1))(v-1) - \cosh(\eta) \sinh(\eta(v-1))\right] = \notag
\\
& = \frac{e^{-\eta}}{4} \int_{-\infty}^\infty dv \ e^{-tv^2/4} v \frac{\sinh(\eta (v - 1))}{\sinh(\eta)} + \frac{e^{-\eta}}{4} \left(2 - \partial_\eta\right) \int_{-\infty}^\infty dv \ e^{-tv^2/4} \frac{\sinh(\eta (v - 1))}{\sinh(\eta)} - \notag
\\
&\hspace{10pt} - \frac{e^{-\eta}}{4} \int_{-\infty}^\infty dv e^{-tv^2/4} \left(v + 2 - \partial_\eta\right) \frac{1}{\sinh(\eta)^2} \left[\frac{v-2}{2v} \sinh(\eta v) - \frac{1}{2} \sinh((v-2)\eta)\right] = \notag
\\
& = \sqrt{\frac{\pi}{t^3}} e^{\eta^2/t} \eta \coth(\eta) e^{-\eta} - \frac{e^{-\eta}}{2} \left(2 - \partial_\eta\right) e^{\eta^2/t} \sqrt{\frac{\pi}{t}} - \notag
\\
&\hspace{10pt} - \frac{e^{-\eta}}{4} \int_{-\infty}^\infty dv e^{-tv^2/4} \left(v + 2\right) \frac{1}{\sinh(\eta)^2} \left[\frac{v-2}{2v} \sinh(\eta v) - \frac{1}{2} \sinh((v-2)\eta)\right] - \notag
\\
&\hspace{10pt} + \frac{e^{-\eta}}{4} \partial_\eta \int_{-\infty}^\infty dv e^{-tv^2/4} \frac{1}{\sinh(\eta)^2} \left[\frac{v-2}{2v} \sinh(\eta v) - \frac{1}{2} \sinh((v-2)\eta)\right] = \notag
\\
& = 2 \sqrt{\frac{\pi}{t^3}} \eta e^{\eta^2/t} e^{-\eta} (1 + \coth(\eta)) \left[1 - \frac{t}{2\eta} \coth(\eta) + \mathcal O(t^2)\right]
\end{align}
Using the fact that $e^{-\eta}(1+\coth(\eta)) = 1/\sinh(\eta)$, this reduces to
\begin{align}
2 \sqrt{\frac{\pi}{t^3}} \frac{\eta e^{\eta^2/t}}{\sinh(\eta)} \left[1 - \frac{t}{2\eta} \coth(\eta) + \mathcal O(t^2)\right]
\end{align}
Thus, putting together with the result above, we find
\begin{align}
\gamma^1_1 & = 2\sqrt{\frac{\pi}{t^3}} \frac{\eta e^{\eta^2/t}}{\sinh(\eta)} \left[1 + \frac{t}{2\eta} \left(\frac{1}{\sinh(\eta)} - \coth(\eta)\right) + \mathcal O(t^2)\right] = \notag
\\
& = 2\sqrt{\frac{\pi}{t^3}} \frac{\eta e^{\eta^2/t}}{\sinh(\eta)} \left[1 - \frac{t}{2\eta} \tanh(\eta/2) + \mathcal O(t^2)\right]
\end{align}
\\
Using that
\begin{align}
\langle 1 \rangle_z = 2\sqrt{\frac{\pi}{t^3}} \frac{\eta e^{\eta^2/t}}{\sinh(\eta)} e^{t/4}
\end{align}
we then conclude that
\begin{align}
\gamma^1_0 = \langle 1 \rangle_z \left[1 + t \frac{2 \sinh(\eta/2)}{\eta \sinh(\eta)} + \mathcal O(t^2)\right]
\end{align}
and
\begin{align}
\gamma^1_1 = \langle 1 \rangle_z \left[1 - t \left(\frac{1}{4} + \frac{1}{2\eta} \tanh(\eta/2)\right) + \mathcal O(t^2)\right]
\end{align}
A similar compuation reveals that
\begin{align}
\gamma^{1/2}_{1/2}&= \langle1\rangle_z \left[1+ \frac{t}{4\eta}\left(\frac{3}{4}\eta-\tanh\left(\frac{\eta}{2}\right)\right) + \mathcal O(t^2)\right]
\end{align}
\end{document}